\documentstyle[epsf]{article}  
\begin{document}
\newcommand {\bea}{\vspace{-.25in}\begin{eqnarray}}   
\newcommand {\eea}{\vspace{-.25in}\end{eqnarray}}


\newcommand {\CD}{\mathrm{C\hspace{.2mm}D}}
\newcommand {\CI}{\mathrm{C\hspace{.2mm}I}}
\newcommand {\n}{\nonumber\\}
\newcommand {\noi}{\noindent}
\newcommand {\p}{{\cal P}}
\newcommand {\la}{\Lambda}
\newcommand {\lll}[1]{\label{eq:#1}}
\newcommand {\reff}[1]{Eq. (\ref{eq:#1})}
\newcommand {\bx}[2]{\Delta^{#1}_{#2}}
\newcommand {\Bx}[2]{\Delta^{#1}_{#2}}
\newcommand {\OR}{{\cal O}}
\newcommand {\SP}{64 \pi^5}
\newcommand {\SPD}{64 \pi^5 p_1^+ \delta^{(5)}(p_1 - p_2 - p_3)}
\newcommand {\Mf}{{\cal M}_f^2}
\newcommand {\M}{{\cal M}^2(\la)}
\newcommand {\V}{V(\la)}
\newcommand {\Vp}{V(\la')}
\newcommand {\Mp}{{\cal M}^2(\la')}
\newcommand {\MI}{{\cal M}^2_I(\la)}
\newcommand {\MIp}{{\cal M}^2_I(\la')}
\newcommand {\lz}{{\la \; \mathrm{terms}}}
\newcommand {\lzb}{\:\rule[-4mm]{.1mm}{8mm}_{\; \la \; \mathrm{terms}}}
\newcommand {\gla}{g_{_{\hspace{-.01in}\la}}}
\newcommand {\glap}{g_{_{\hspace{-.02in}\la'}}}
\newcommand {\rsq}{\vec r^{\, 2}_\perp}
\newcommand {\qsq}{\vec q^{\, 2}_\perp}
\newcommand {\ksq}{\vec k^{\, 2}_\perp}
\newcommand {\wsq}{\vec w^{\, 2}_\perp}
\newcommand {\D}[2]{\, \delta_{#1,#2} \, }
\newcommand {\Hf}{H_\mathrm{free}}
\newcommand {\Hi}{H_\mathrm{int}}
\newcommand {\dVme}{\left< F \right| \delta V \left| I \right>}
\newcommand {\dV}{\delta V}
\newcommand {\dVo}[1]{\delta V^{(#1)}}
\newcommand {\dVome}[1]{\left< F \right| \delta V^{(#1)} \left| I \right>}
\newcommand {\dVotwototwo}[1]{\left< \phi_3 \phi_4 \right| \delta V^{(#1)} \left| \phi_1 \phi_2 \right>}
\newcommand {\dVoonetothree}[1]{\left< \phi_2 \phi_3 \phi_4 \right| \delta V^{(#1)} \left| \phi_1 \right>}
\newcommand {\dVthreeonetotwo}{\left< \phi_2 \phi_3 \right| \delta V^{(3)} \left| \phi_1 \right>}
\newcommand {\Vthreeonetotwo}{\left< \phi_2 \phi_3 \right| V^{(3)}_{\CD}(\la) \left| \phi_1 \right>}

\pagestyle{empty}

{\Large \bf \centerline{Systematic Renormalization in Hamiltonian}}
{\Large \bf \centerline{Light-Front Field Theory}}

\vspace{.2in}

\centerline{Brent H. Allen\footnote{E-mail: allen@mps.ohio-state.edu} and Robert J. Perry\footnote{E-mail:
perry@mps.ohio-state.edu}}

\vspace{.1in}

\centerline{\it Department of Physics, Ohio State University, Columbus, Ohio 43210}

\centerline{(April 1998)}
\vspace{.5in}

\centerline{\bf Abstract}

\vspace{-.1in}

\begin{quotation}
We develop a systematic method for computing a renormalized light-front field theory Hamiltonian that can
lead to bound states that rapidly converge in an expansion in free-particle Fock-space sectors. To
accomplish this without dropping any Fock states from the theory, and to regulate the Hamiltonian, we
suppress the matrix elements of the Hamiltonian between free-particle Fock-space states that differ in free
mass by more than a cutoff. The cutoff violates a number of physical principles of the theory, and thus the
Hamiltonian is not  just the canonical Hamiltonian with masses and couplings redefined by renormalization. 
Instead, the Hamiltonian must be allowed to contain all operators that are consistent with the unviolated
physical principles of the  theory.  We show that if we require the Hamiltonian to produce
cutoff-independent physical quantities and we require it to respect the unviolated physical principles of
the theory, then its matrix elements are uniquely determined in terms of the fundamental parameters of the
theory. This method is designed to be applied to QCD, but, for simplicity, we illustrate our method by
computing and analyzing second- and third-order matrix elements of the Hamiltonian in massless $\phi^3$
theory  in six dimensions.

\vspace{.1in}
\noi
PACS number(s): 11.10.Gh
\end{quotation}

\newpage
\pagestyle{plain}

\setcounter{footnote}{0}

\section{Introduction}

It may be possible for hadron states to rapidly converge in an expansion in free-particle Fock-space sectors
(free sectors) in Hamiltonian light-front quantum chromodynamics (HLFQCD).  This will happen if the Hamiltonian
satisfies three conditions.  First, the diagonal matrix elements of the Hamiltonian in the basis of free-particle
Fock-space states (free states) must be dominated by the free part of the Hamiltonian.   Second, the off-diagonal
matrix elements of the Hamiltonian in this basis must quickly decrease as the difference of the free masses of
the states increases.  If the Hamiltonian satisfies these first two conditions, then each  of its eigenstates
will be dominated by free-state components with free masses that are close to the mass of the eigenstate.  
The third
condition on the Hamiltonian is that the free mass of a free state must quickly increase as the number of 
particles in
the state increases.  If the Hamiltonian satisfies all three conditions\footnote{Exactly how quickly the
Hamiltonian's off-diagonal matrix elements must decrease and the mass of a free state must increase  is not
known; so we assume that the rates that we are able to achieve are sufficient. This must be verified by
diagonalizing the Hamiltonian.}, then the number of particles in a free-state component 
that dominates an eigenstate  will
be limited from above by the constraint that the free mass of the free-state component must be close to the mass of the
eigenstate. This means that the  eigenstate will rapidly converge in an expansion in free sectors\footnote{The
coefficients of the expansion for highly excited eigenstates may grow for a number of free sectors and then peak
before diminishing and becoming rapidly convergent.}.

We argue that if one uses Hamiltonian light-front field theory (HLFFT) rather than equal-time field theory, then
it is possible to calculate a Hamiltonian that satisfies these conditions.  If the field theory is to be
complete, then the Hamiltonian must satisfy these conditions without a truncation of the Fock space or the
removal of particle-number-changing interactions.  Based on the early work of
Dyson [1] and Wilson [2], and the more recent work of Wegner [3] and G{\l}azek and Wilson [4], there have been a
substantial number of efforts to derive such Hamiltonians perturbatively [5-12].  However, a calculation of the
QCD Hamiltonian beyond second order in perturbation theory requires one to take the scale dependence of the
coupling into account, and none of these earlier methods demonstrates how to do this.  In this paper
we develop an alternative method for calculating a renormalized Hamiltonian that satisfies the conditions
discussed above.  We do not truncate the Fock space or remove particle-number-changing interactions.  We show 
how the scale dependence of the coupling is taken into account in general and we present explicit examples that 
demonstrate how this is done in practice. 

If we are to be able to derive a Hamiltonian that satisfies the conditions discussed above, then we must work
in an approach in which it is possible for the vacuum to be dominated by 
few-body free sectors.  This does not seem possible in equal-time field theory unless 
the volume of space is severely limited.  In HLFFT, however, each particle in the
vacuum  has a longitudinal momentum of exactly zero\footnote{This is because there are no negative longitudinal 
momenta and momentum conservation requires the three-momenta of the constituents of the vacuum to  sum to
zero.}.  This means that we can force the vacuum to be empty by requiring every particle to have a positive 
longitudinal momentum. The effects of the excluded particles must be replaced by interactions in the Hamiltonian,
and a calculation of these interactions may have to be nonperturbative.   We view the calculation of these
interactions as an extension of our method and plan to include them in the future.  Until we do consider these
interactions, we are avoiding the vacuum problem rather than solving it.

In light-front field theory, all the dynamics of the Hamiltonian are in the invariant-mass operator and their
matrix elements are trivially related.  For convenience, we work with the invariant-mass operator directly  and
refer to it as the Hamiltonian.  To see how we can derive a Hamiltonian that satisfies our conditions, we  divide
it into a free part and an interacting part:

\bea
\M = \Mf + \MI.  
\eea

\noi
The interacting part of the Hamiltonian is a function of the regulator, $\la$.
The free states are eigenstates of the free Hamiltonian:

\bea
\Mf \left| F \right> = M_F^2 \left| F \right> .
\eea

\noi
The diagonal matrix elements of the Hamiltonian are given by

\bea
\left< F \right| \M \left| F \right> &=& \left< F \right| \Mf \left| F
\right> + 
\left< F \right| \MI \left| F \right> \n
&=& M_F^2 \left<F \right| \left. \! \! F \right> +
\left< F \right| \MI \left| F \right> .
\eea

\noi
We assume that the Hamiltonian can be computed perturbatively, in which case the free part of the Hamiltonian will
dominate the diagonal matrix elements.  This fulfills the first condition on the Hamiltonian.

We regulate the Hamiltonian with a cutoff $\la$ such that

\bea
\left< F \right| \M \left| I \right> \sim e^{\frac{-(M_F^2 - M_{I}^2)^2}{\Lambda^4}} ,
\lll{gauss}
\eea

\noi
and then as $\left|M_F^2 - M_{I}^2 \right|$ grows, the off-diagonal matrix elements of the Hamiltonian quickly
diminish,  thus fulfilling the second condition on the Hamiltonian. We show in Appendix A that in HLFQCD it is
reasonable to expect that $M_F^2$ grows as least quadratically with the number of particles in $\left| F
\right>$, perhaps much faster.  This fulfills the third and final condition on the Hamiltonian.

The cutoff violates a number of physical principles of the theory, such as Lorentz invariance and gauge
invariance.  This implies that the Hamiltonian cannot be  just the canonical Hamiltonian with masses and
couplings redefined by renormalization.  Instead, the Hamiltonian must be allowed to contain all operators that
are consistent with the unviolated physical principles of the  theory.  The key point of this paper is that if
we require the Hamiltonian to produce cutoff-independent physical quantities and we require it to respect the
unviolated physical principles of the theory, then its matrix elements are uniquely
determined in terms of the fundamental parameters of the theory.  Although the focus of this paper is the
computation of the Hamiltonian, any observable can be calculated in our approach.

Our main assumption is that in an asymptotically free theory the Hamiltonian can be determined 
perturbatively, provided the cutoff is chosen so that the couplings are sufficiently small.
Although we compute the Hamiltonian perturbatively, it can be used to obtain nonperturbative quantities.  For
example, bound-state masses can be computed by diagonalizing the Hamiltonian. The disadvantage of perturbative
renormalization is that nonperturbative physical quantities will be somewhat cutoff-dependent.  The strength of
this cutoff dependence has to be checked after nonperturbative physical quantities are computed and is not
considered here.

In general, field theories have an infinite number of degrees of freedom. However, since our Hamiltonian will
cause hadron states to rapidly converge in an expansion in free sectors, approximate computations of physical
quantities will require only finite-body matrix elements of the Hamiltonian.  In addition, since we assume
that we can renormalize the Hamiltonian perturbatively, we do not implement particle creation and annihilation
nonperturbatively in the renormalization process.  This allows us to work with a finite number of degrees of
freedom when calculating the required finite-body matrix elements of the Hamiltonian.

We are going to apply the method to QCD
in future work; however, the complexities of QCD are largely irrelevant to its development.  For this reason, we
choose to illustrate the method with a much simpler theory, massless $\phi^3$ theory in six dimensions.  This
theory has some similarities to QCD that make it  a suitable testing ground.  In particular, its diagrams have a
similar structure to QCD  diagrams and it is asymptotically free.  Nonperturbatively the theory is unstable;
however, we only use it to illustrate how to compute the Hamiltonian perturbatively, and so the instability is
irrelevant.

In Section 2 we specify how the Hamiltonian is regulated and restrict it to produce cutoff-independent physical
quantities. In Section 3 we restrict the Hamiltonian to respect the physical principles of the theory that are
not violated by the cutoff.    In Section 4 we show how the restrictions that we place on the Hamiltonian allow
us to  uniquely determine its matrix elements in terms of the fundamental parameters of the theory.  Section 5
contains example calculations of a few  second-order matrix elements of the Hamiltonian.  Section 6 contains a
perturbative example of how the cut-off Hamiltonian fulfills the restriction that it must lead to
cutoff-independent physical quantities. In Section 7 we calculate the matrix element of the Hamiltonian for $\phi
\rightarrow \phi \phi$ through third order\footnote{We neglect  contributions to this matrix element that depend
on the cutoff only through the Gaussian regulating factor.   See Sections 4 and 7 for details.} and demonstrate
how the coupling runs in our approach.   Finally, in Section 8 we provide a summary and a short discussion of the
extension of our method to QCD.

\section{Regulation and Renormalization}

\subsection{The Cutoff}

We work in the basis of free states.  We write the Hamiltonian as the sum of 
the canonical free part and an interaction (see Appendix B for conventions and definitions):

\bea
\M = \Mf + \MI .
\lll{H split}
\eea

\noi
At this point the interaction is undefined.  To force the off-diagonal matrix
elements of the Hamiltonian to decrease quickly for increasingly different free masses, we regulate the
Hamiltonian by suppressing its matrix elements between states that differ in free mass by more than a cutoff:

\bea
\left< F \right| \M \left| I \right> &=& \left< F \right| \Mf \left| I \right> + 
\left< F \right| \MI \left| I \right> \n
&=& M_F^2 \left< F \right| \!I \left. \!\right> + 
e^{-\frac{\bx{2}{FI}}{\la^4}} \left<F \right| \V \left| I \right> ,
\lll{cutoff}
\eea

\noi
where $\left| F \right>$ and $\left| I \right>$ are eigenstates of the free Hamiltonian
with eigenvalues $M_F^2$ and $M_I^2$, and $\Bx{}{FI}$ is the difference of these eigenvalues:

\bea
\Bx{}{FI} = M_F^2 - M_I^2 .
\eea

\noi
$\V$ is the interaction with the Gaussian factor removed, and we refer to it as the ``reduced interaction."
To determine the Hamiltonian, we must determine the reduced interaction.  

Assuming $\left<F \right| \V \left| I \right>$ does not grow exponentially as $\Bx{2}{FI}$ gets large,  the
exponential in \reff{cutoff} suppresses  each off-diagonal matrix element of the Hamiltonian for which
$\Bx{2}{FI}$ is large compared to $\la^4$.   This regulates the Hamiltonian and forces it to weakly couple free
states with significantly different free masses.

\subsection{The Restriction to Produce Cutoff-Independent Physical Quantities}

Unfortunately, any regulator in HLFFT breaks Lorentz  invariance, and in gauge theories it also breaks
gauge invariance\footnote{Our regulator breaks these symmetries because the mass of a free state is neither
gauge-invariant nor rotationally invariant (except for transverse rotations).}.  This
means that in HLFFT
renormalization cannot be  performed simply by redefining the canonical couplings and masses.  Instead, the Hamiltonian
must be allowed to contain all operators that are consistent with the unviolated physical principles of the
theory.  In addition, since there is no locality in the longitudinal direction in HLFFT\footnote{That there is no
longitudinal locality in HLFFT is evident from the fact that the longitudinal momentum of a free particle appears
in
the denominator of its dispersion relation, $(\vec p_\perp^{\,2} + m^2)/p^+$.}, these operators can contain
arbitrary functions of longitudinal momenta.

To uniquely determine the operators that the Hamiltonian can contain, as well as the coefficients of these
operators,
we place a number of restrictions on the Hamiltonian.  The first restriction is that the Hamiltonian has to
produce
cutoff-independent physical quantities.  We impose this restriction by requiring the Hamiltonian to satisfy

\bea
\M = U(\la,c \la) \: {\cal M}^2(c \la) \: U^\dagger(\la,c \la) ,
\lll{unitary equiv1}
\eea

\noi
where $c > 1$ and is otherwise arbitrary, and $U$ is a unitary transformation that changes the Hamiltonian's
cutoff. Note that we are considering ${\cal M}^2$ to be a function
of its argument; i.e. ${\cal M}^2(c \la)$ has the same functional dependence on $c \la$ that $\M$ has on
$\la$.  To see that \reff{unitary equiv1} implies that the Hamiltonian will produce cutoff-independent physical quantities, note
that the Hamiltonian on the right-hand side (RHS) of the equation can be replaced by iterating
the equation:

\bea
\M &=& U(\la,c \la) \: {\cal M}^2(c \la) \: U^\dagger(\la,c \la) \n
&=& \Big[ U(\la,c \la) \: U(c \la,c^2 \la) \Big] \: {\cal M}^2(c^2 \la) \: \Big[U^\dagger(c \la,c^2 \la) \:
U^\dagger(\la,c \la)\Big] \n
&=& \Big[U(\la,c \la) \: U(c \la,c^2 \la) \: U(c^2 \la, c^3 \la)\Big] \: {\cal M}^2(c^3 \la) \:
\Big[U^\dagger(c^2 \la, c^3 \la) \: U^\dagger(c \la,c^2 \la) \: U^\dagger(\la,c \la)\Big] \n
&\vdots& \hspace{.5in}.
\eea

\noi
Thus, since $c > 1$, \reff{unitary equiv1} implies that $\M$ is unitarily equivalent to 
$\lim_{\la \rightarrow \infty} \M$.
This means $\M$ will produce cutoff-independent physical quantities\footnote{If the theory is not
asymptotically free, then this argument cannot be used in perturbation theory.}.

For simplicity, we define

\bea
\la' = c \la ,
\lll{lap1}
\eea

\noi
and then \reff{unitary equiv1} takes the form

\bea
\M = U(\la,\la') \: {\cal M}^2(\la') \: U^\dagger(\la,\la') .
\lll{unitary equiv}
\eea

To calculate the Hamiltonian perturbatively, we need to be able to implement this
equation perturbatively.  In our approach, the cutoff that regulates a Hamiltonian also serves to define the
scale of the coupling that appears in the Hamiltonian; so the coupling in $\M$ is defined at the
scale $\la$.  To use \reff{unitary equiv} perturbatively,
we have to perturbatively relate the coupling at the scale $\la$ to the coupling at the scale
$\la'$.  This perturbative relationship is well-defined only if $\la'$ is not very large compared
to $\la$ [13].  To fulfill this requirement and \reff{lap1} (recall $c > 1$), we choose $\la'$ to satisfy

\bea
\la < \la' < 2 \la .
\lll{lap}
\eea

\noi
Now \reff{unitary equiv}, which restricts the Hamiltonian to produce
cutoff-independent physical quantities, can be imposed perturbatively.
 
\subsection{The Unitary Transformation}

To make \reff{unitary equiv} a complete statement, we have to define the unitary transformation.  
The transformation is designed to alter the cutoff implemented in \reff{cutoff}, and is a simplified 
version of a transformation introduced by Wegner [3], modified for
implementation with the invariant-mass operator.  The transformation is uniquely defined by a linear
first-order differential equation:

\bea
\frac{d U(\la,\la')}{d (\la^{-4})} = T(\la) U(\la,\la') ,
\lll{diff eq}
\eea

\noi
with one boundary condition:

\bea
U(\la,\la) = {\bf 1} .
\eea

\noi
It is shown in Appendix C that $U(\la,\la')$ is unitary as long as $T(\la)$ is anti-Hermitian.  We define

\bea
T(\la) = \left[ \Mf, \M \right] ,
\lll{T def}
\eea

\noi
which is anti-Hermitian.  The transformation introduced by Wegner uses

\bea
T(\la) = \left[ {\cal M}_{\mathrm{D}}^2(\la), \M \right] ,
\eea

\noi
where ${\cal M}_{\mathrm{D}}^2(\la)$ is the part of $\M$ that is diagonal in the free basis. (It includes the diagonal
parts
of all interactions.)  Our transformation regulates
changes in eigenvalues of $\Mf$, and Wegner's transformation regulates
changes in eigenvalues of ${\cal M}_{\mathrm{D}}^2(\la)$.

To solve for $\M$ perturbatively, we need to turn \reff{unitary equiv} into a perturbative restriction on
the reduced interaction, $\V$.  We begin by taking the derivative of \reff{unitary equiv}:

\bea
\frac{d \M}{d (\la^{-4})} &=& \left[ \left[ \Mf, \M \right], \M \right] ,
\lll{ham com}
\eea

\noi
where \reff{diff eq} and its adjoint and \reff{T def} are used\footnote{The restrictions that we place on
the Hamiltonian in the next section prohibit $\Mp$ from depending on $\la$.}.  Since the free part of the
Hamiltonian commutes with itself and is independent of the cutoff,

\bea
\frac{d \MI}{d (\la^{-4})} = \left[ \left[ {\cal M}_f^2, \MI \right], \M \right] .
\lll{M diff}
\eea

We assume that $\M$ can be written as a convergent (or at least asymptotic) expansion in powers of $\MIp$:

\bea
\M = \sum_{a=0}^\infty v^{(a)} ,
\lll{int exp}
\eea

\noi
where $v^{(a)}$ is $\OR\left(\left[\MIp \right]^a\right)$, $v^{(0)}=\Mf$, and we must determine
the higher-order $v^{(a)}$'s.
Substituting \reff{int exp} into \reff{M diff}, matching powers of $\MIp$ on both sides, and doing some 
algebra yields

\bea
\frac{d v^{(a)}}{d (\la^{-4})} = \sum_{b=1}^a \left[ \left[ \Mf , v^{(b)} \right] , v^{(a-b)} 
\right] ,
\lll{m diff}
\eea

\noi
where the sum is zero if $a=0$.
This equation tells us how the $\OR\left(\left[\MIp \right]^a\right)$ contribution to $\M$ depends on the cutoff
in terms of lower-order
contributions, and can be used to iteratively calculate $\M$ in powers of $\MIp$.

Taking a matrix element of \reff{m diff} between free states $\left| i \right>$ and $\left| j \right>$ for the
case 
$a=1$, we obtain

\bea
\frac{d v_{ij}^{(1)}}{d (\la^{-4})} = - v_{ij}^{(1)} \, \Bx{2}{ij} .
\eea

\noi
Solving this equation with the boundary condition

\bea
\M \:\rule[-3mm]{.1mm}{6mm}_{\; \la \rightarrow \la'} = \Mp 
\lll{bc}
\eea

\noi
leads to

\bea
v_{ij}^{(1)} = \exp \left[ \left( \frac{1}{\la'^4} - \frac{1}{\la^{4}} \right) \Bx{2}{ij} \right] \MIp_{ij} .
\lll{linear}
\eea

\noi
$\MIp$ is regulated such that $\MIp_{ij}$ is proportional to $\exp (-\bx{2}{ij} \la'^{-4})$, which means that
\reff{linear} shows that
$v_{ij}^{(1)}$ is proportional to $\exp (-\bx{2}{ij} \la^{-4})$.  

The higher-order $v^{(a)}$'s can be found the
same
way that $v^{(1)}$ was found, by taking matrix elements of \reff{m diff} and using the lower-order $v^{(a)}$'s
and
the boundary condition, \reff{bc}.  When products of $\MIp$ are encountered, complete sets of free states have to
be inserted between adjacent factors.  The $v^{(a)}$'s that result from this procedure are all proportional to 
$\exp (-\bx{2}{ij} \la^{-4})$, which 
shows that $U(\la,\la')$ does indeed take a Hamiltonian with a cutoff $\la'$ and produce one with a cutoff
$\la$.

If one computes $v^{(2)}$ and $v^{(3)}$, then \reff{int exp} gives $\M$ in terms of $\MIp$ to 
$\OR\left(\left[\MIp \right]^3\right)$.  Using this and 

\bea
\left<F \right| \MI \left| I \right> = 
e^{-\frac{\bx{2}{FI}}{\la^4}} \left<F \right| \V \left| I \right> ,
\eea

\noi
which follows from \reff{cutoff}, one can show that the perturbative version of \reff{unitary equiv} in terms
of the reduced interaction is

\bea
\V - \Vp &=& \dV ,
\lll{unitary const}
\eea

\noi
where $\dV$ is the change in the reduced interaction and is a function of both $\la$ and $\la'$:

\bea
\dVme &=& \frac{1}{2} \sum_K
\left<F \right| \Vp \left| K \right> \left<K \right| \Vp \left| I \right> T_2^{(\la,\la')}(F,K,I) \n
&+& \frac{1}{4} \sum_{K,L}  \left<F \right| \Vp \left| K \right> \left<K \right| \Vp \left| L \right>
\left<L \right| \Vp \left| I \right> T_3^{(\la,\la')}(F,K,L,I) \n
&+& \OR\left(\left[\Vp \right]^4\right) .
\lll{RFI}
\eea

\noi
In this equation, the sums are over complete sets of free states and the cutoff functions are defined by

\bea
T_2^{(\la,\la')}(F,K,I) &=& \left( \frac{1}{\Bx{}{FK}} - \frac{1}{\Bx{}{KI}} \right) \left( 
e^{2 \la'^{-4} \bx{}{FK} \bx{}{KI}} - e^{2 \la^{-4} \bx{}{FK} \bx{}{KI}} \right)
\eea

\noi
and

\bea
&&T_3^{(\la,\la')}(F,K,L,I) = \n
&&\left( \frac{1}{\Bx{}{KL}} - \frac{1}{\Bx{}{LI}} \right) \left( \frac{1}{\Bx{}{KI}} -
\frac{1}{\Bx{}{FK}} \right) e^{2 \la'^{-4} \bx{}{KL} \bx{}{LI}} \left( e^{2 \la^{-4} \bx{}{FK} \bx{}{KI}} -
e^{2 \la'^{-4} \bx{}{FK} \bx{}{KI}} \right) \n
&+& \left( \frac{1}{\Bx{}{KL}} - \frac{1}{\Bx{}{LI}} \right) 
\frac{\Bx{}{FK} + \Bx{}{IK}}{\Bx{}{KL} \Bx{}{LI} + \Bx{}{FK} \Bx{}{KI}} \left(  e^{2 \la'^{-4} (\bx{}{FK}
\bx{}{KI}
+ \bx{}{KL} \bx{}{LI} )} - 
e^{2 \la^{-4} (\bx{}{FK} \bx{}{KI} + \bx{}{KL} \bx{}{LI} )} \right) \n
&+& \left( \frac{1}{\Bx{}{FK}} -
\frac{1}{\Bx{}{KL}} \right) \left( \frac{1}{\Bx{}{LI}} -
\frac{1}{\Bx{}{FL}} \right) e^{2 \la'^{-4} \bx{}{FK} \bx{}{KL}} \left( e^{2 \la^{-4} \bx{}{FL} \bx{}{LI}} -
e^{2 \la'^{-4} \bx{}{FL} \bx{}{LI}} \right) \n
&+& \left( \frac{1}{\Bx{}{FK}} - \frac{1}{\Bx{}{KL}} \right) 
\frac{\Bx{}{FL} + \Bx{}{IL}}{\Bx{}{FK} \Bx{}{KL} + \Bx{}{FL} \Bx{}{LI}}  \left(  e^{2 \la'^{-4} (\bx{}{FK}
\bx{}{KL} + \bx{}{FL} \bx{}{LI} )} - 
e^{2 \la^{-4} (\bx{}{FK} \bx{}{KL} + \bx{}{FL} \bx{}{LI} )}  \right) . \n
\eea

\noi
The above definitions for the cutoff functions assume that none of the $\Bx{}{}$'s that appear
in the denominators is zero.  In the event one of them is zero, the appropriate cutoff function is defined by

\bea
T^{(\la,\la')}_i(\bx{}{}=0) &=& \lim_{\bx{}{} \rightarrow 0} T^{(\la,\la')}_i(\bx{}{}) .
\eea

To summarize,
\reff{unitary const} is the perturbative version of the restriction that $\M$ has to produce cutoff-independent 
physical quantities, expressed in terms of the reduced interaction and the change in the reduced interaction.

\section{Restrictions on the Hamiltonian from Physical Principles}

\reff{unitary const} is the first restriction on the Hamiltonian.
To uniquely determine the Hamiltonian, we need to place additional restrictions on it, and we do this using 
the physical principles of the theory that are not violated by the cutoff.

\subsection{Symmetry Properties}

Any LFFT should exhibit momentum conservation, boost invariance, and transverse rotational invariance.  
Our cutoff does not violate any
of these principles; so we restrict the Hamiltonian to conserve momentum and to be invariant under boosts
and transverse rotations.

\subsection{Cluster Decomposition}

Cluster decomposition is a physical principle of LFFT that is partially violated by our cutoff.
However, we can identify the source of the violation and still use the principle to restrict the 
Hamiltonian.  

To see how to do this, note that momentum conservation implies that any matrix 
element of the reduced interaction can be written as a sum of terms, 
with each term containing a unique product of
momentum-conserving delta functions.  For example, 
$\left< \phi_3 \phi_4 \right| V(\la) \left| \phi_1 \phi_2 \right>$ 
can be written in the form

\bea
\left< \phi_3 \phi_4 \right| V(\la) \left| \phi_1 \phi_2 \right> &=& \left[ \SP (p_1^+ + p_2^+) 
\delta^{(5)}(p_1 + p_2 - p_3 - p_4) \right] F^{(1)}(p_1,p_2,p_3,p_4,\la) \n
&+& \left[ \SP p_1^+ \delta^{(5)}(p_1 - p_3) \right] \left[ \SP p_2^+ \delta^{(5)}(p_2 - p_4) \right] 
F^{(2)}(p_1,p_2,p_3,p_4,\la) \n
&+&  \left[ \SP p_1^+ \delta^{(5)}(p_1 - p_4) \right] \left[ \SP p_2^+ \delta^{(5)}(p_2 - p_3) \right] 
F^{(3)}(p_1,p_2,p_3,p_4,\la) .
\lll{ex delta exp}
\eea

\noi
We have included a factor of $\SP$ and a longitudinal momentum factor with each
delta function because our normalization of states produces these factors naturally.
Cluster decomposition implies that the $F^{(i)}$'s in \reff{ex delta exp} cannot contain
delta functions of momenta [15].  It also implies that each 
term in \reff{ex delta exp} has exactly one delta function associated with the conservation of momenta of
interacting
particles, with any additional delta functions associated with the conservation of momenta of spectators.
This means that the first term on the RHS of \reff{ex delta exp} is the contribution from four interacting 
particles, and the second term on the RHS can have two contributions: one for which
$\phi_1$ and $\phi_3$ are spectators, and one for which $\phi_2$ and $\phi_4$ are spectators.  Similarly,
the third term on the RHS can have two contributions: one for which
$\phi_1$ and $\phi_4$ are spectators, and one for which $\phi_2$ and $\phi_3$ are spectators.  

Normally, contributions to matrix elements depend on spectators only through momentum-conserving delta functions
and the corresponding factors of longitudinal momenta, in which case we can write

\bea
&&\left< \phi_3 \phi_4 \right| V(\la) \left| \phi_1 \phi_2 \right> = \left[ \SP (p_1^+ + p_2^+) 
\delta^{(5)}(p_1 + p_2 - p_3 - p_4) \right] 
F^{(1)}(p_1,p_2,p_3,p_4,\la) \n
&+& \left[ \SP p_1^+ \delta^{(5)}(p_1 - p_3) \right] \left[ \SP p_2^+ \delta^{(5)}(p_2 - p_4) \right] 
\left[ F^{(2,1)} (p_2,p_4,\la) 
+ F^{(2,2)} (p_1,p_3,\la) \right] \n
&+&  \left[ \SP p_1^+ \delta^{(5)}(p_1 - p_4) \right] \left[ \SP p_2^+ \delta^{(5)}(p_2 - p_3) \right] 
\left[ F^{(3,1)} (p_2,p_3,\la) 
+ F^{(3,2)} (p_1,p_4,\la) \right] ,
\lll{cluster 1}
\eea

\noi
where we explicitly show the contributions from different sets of spectators. 
However, we use a cutoff on differences of free masses of states, and the free mass of a state does not
separate into independent contributions from each particle in the state.  A particle's contribution to the
free mass of a state is $\vec r_\perp^{\,2}/x$, where $x$ is the fraction of the state's 
longitudinal momentum carried by the particle and $\vec r_\perp$ is the transverse momentum of the particle
in the state's center-of-mass frame.  When the longitudinal momentum of any other particle in the state changes, 
$x$ changes because the total longitudinal momentum of the state changes.
Thus our cutoff partially violates the cluster decomposition principle, such that
the $F^{(i,j)}$'s in \reff{cluster 1} can depend on $P^+$, the total longitudinal momentum of the states:

\bea
&&\left< \phi_3 \phi_4 \right| V(\la) \left| \phi_1 \phi_2 \right> = \left[ \SP (p_1^+ + p_2^+) 
\delta^{(5)}(p_1 + p_2 - p_3 - p_4) \right] 
F^{(1)}(p_1,p_2,p_3,p_4,\la) \n
&+& \left[ \SP p_1^+ \delta^{(5)}(p_1 - p_3) \right] \left[ \SP p_2^+ \delta^{(5)}(p_2 - p_4) \right] 
\left[ F^{(2,1)}(p_2,p_4,P^+,\la) + F^{(2,2)} (p_1,p_3,P^+,\la) \right] \n
&+&  \left[ \SP p_1^+ \delta^{(5)}(p_1 - p_4) \right] \left[ \SP p_2^+ \delta^{(5)}(p_2 - p_3) \right] 
\left[ F^{(3,1)} (p_2,p_3,P^+,\la) + F^{(3,2)} (p_1,p_4,P^+,\la) \right] \!\!.
\lll{cluster 2}
\eea

\noi
The result of this analysis can be generalized to formulate a restriction on the Hamiltonian:
when a matrix element of the reduced interaction
is written as an expansion in the possible combinations of momentum-conserving
delta functions, the coefficient of any term in the expansion depends only on the momenta of the particles that
are interacting in the term, the total longitudinal momentum of the states, and the cutoff.

It is worth mentioning that if one uses a cutoff on free masses  of states rather than differences of free
masses,
then the $F^{(i,j)}$'s in  \reff{cluster 2} can be functions of all the momenta in the matrix element, 
which is a signal of a much more
severe violation of cluster decomposition.  If one uses a cutoff on free energy differences, then cluster
decomposition  is not violated at all, but longitudinal boost invariance is lost.

\subsection{Transverse Locality}

Ideally, a LFFT Hamiltonian should be local in the transverse directions, and thus each of its matrix elements should be
expressible as a finite series of powers of transverse momenta with expansion coefficients that are functions of
longitudinal momenta\footnote{This series is actually multiplied by a product of momentum-conserving delta
functions which has no impact on the locality of the Hamiltonian.}.  In our case, the cutoff suppresses
interactions that have large transverse momentum transfers and replaces them with interactions that have smaller
transverse momentum transfers.  This is equivalent to suppressing interactions that occur over small transverse
separations and replacing them with interactions that occur over larger transverse separations; so we do not
expect our interactions to be perfectly transverse-local.  Nonetheless, we expect that interactions in $\M$
should appear local relative to transverse separations larger than $\la^{-1}$ or, equivalently, to transverse
momenta less than $\la$.  This means that for transverse momenta less than $\la$ we should be able to
approximate each matrix element of $\M$ as a finite power series in $\vec p_\perp/\la$. We enforce this
by assuming that transverse locality is violated in the weakest manner possible, i.e. that  any matrix
element of the Hamiltonian  can be expressed as an {\it infinite} series of powers of transverse momenta with an
infinite radius of convergence. 

\subsection{Representation of the Theory of Interest}

In the remainder of this section, we place a number of restrictions on the Hamiltonian that limit it to describing 
the particular LFFT of interest, massless $\phi^3$ theory.  We work in six dimensions so that the theory is
asymptotically free.

We assume that we can compute the Hamiltonian
perturbatively, which means that we can expand $\V$ in powers of the coupling  at
the scale $\la$.  Our cutoff has no effect in the noninteracting limit; so our Hamiltonian must
reproduce free massless scalar field theory in this limit.  According to \reff{cutoff}, this means that 
$\V$ vanishes in the noninteracting limit.

In massless $\phi^3$ theory, the
only fundamental parameter is the coupling; so we require the Hamiltonian to depend only on it and
the scale.  In this case, the expansion of $\V$ takes the form

\bea
\V = \sum_{r=1}^\infty \gla^r V^{(r)}(\la) ,
\lll{order exp}
\eea

\noi
where $\gla$ is the coupling at the scale $\la$.  We refer to $V^{(r)}(\la)$ as the $\OR(\gla^r)$ reduced
interaction, although for convenience the coupling is factored out.

$\gla$ is the correct fundamental parameter for massless $\phi^3$ theory if and only if
its definition is consistent with the canonical definition of the coupling.  The canonical definition is

\bea
g = \left[ \SP p_1^+ \delta^{(5)}(p_1 - p_2 - p_3) \right]^{-1} \left< \phi_2 \phi_3 \right| 
{\cal M}^2_{\mathrm{can}}
\left| \phi_1 \right> .
\lll{can g}
\eea

\noi
We choose

\bea
\gla &=& \left[ \SP p_1^+ \delta^{(5)}(p_1 - p_2 - p_3) \right]^{-1} \left< \phi_2 \phi_3 \right| \M
\left| \phi_1 \right> \:\rule[-3mm]{.1mm}{6mm}_{\; \vec p_{2 \perp} = \vec p_{3 \perp} = 0 \, ; \, 
p_2^+ = p_3^+ = \frac{1}{2} p_1^+} \n
&=& \left[ \SP p_1^+ \delta^{(5)}(p_1 - p_2 - p_3) \right]^{-1} \left< \phi_2 \phi_3 \right| \V
\left| \phi_1 \right> \:\rule[-3mm]{.1mm}{6mm}_{\; \vec p_{2 \perp} = \vec p_{3 \perp} = 0 \, ; \, 
p_2^+ = p_3^+ = \frac{1}{2} p_1^+} .
\lll{g def}
\eea

\noi
Our definition of the coupling is consistent with the canonical definition
because the conditions on the matrix elements in \reff{g def} have no effect on the matrix element in
\reff{can g}, which is momentum-independent.  
According to \reff{order exp}, the Hamiltonian is {\it coupling coherent} [14] 
because the couplings of its noncanonical operators are functions only of the fundamental parameters 
of the theory and they vanish in the noninteracting limit.

If the Hamiltonian is to produce the correct second-order scattering amplitudes for massless $\phi^3$ theory,
then the first-order reduced interaction must be the canonical interaction:

\bea
V^{(1)} &=& (2 \pi)^5 \p^+ \int D_1 D_2 D_3 \left[ a_3^\dagger a_1 a_2 
\delta^{(5)}(p_3 - p_1 - p_2) + a_2^\dagger a_3^\dagger a_1 
\delta^{(5)}(p_2 + p_3 - p_1) \right] .
\lll{O1}
\eea

\noi
A proof of this statement exists, but the proof is tedious, and so we neglect to present it.

A perturbative scattering amplitude in $\phi^3$ theory depends only on odd powers of the coupling
if the number of particles changes by an odd number and only on even powers of the coupling
if the number of particles changes by an even number.  We require our Hamiltonian to produce
perturbative scattering amplitudes with this feature.

In the remaining sections, we show how the restrictions that we have placed on the Hamiltonian uniquely determine
it in terms of the fundamental parameters of the theory and allow it to produce correct physical quantities.  
As a check on our procedure, we can verify that the
Hamiltonian that results from the procedure satisfies the restrictions that we have placed on it. The
restrictions that we use are based on a subset of the physical principles of the theory.  The remaining
principles, such as Lorentz invariance and gauge invariance (in a gauge theory), must be automatically 
respected by physical quantities derived from
our Hamiltonian, at least perturbatively.  If they are not, then they contradict the principles we use and no
consistent theory can be built upon the complete set of principles.

\section{The Method for Computing Matrix Elements of the Hamiltonian}

In this section we develop a general formalism for using the restrictions that we have placed on the Hamiltonian
to calculate its matrix elements.  To begin, we consider the restriction that
forces the Hamiltonian to produce cutoff-independent physical quantities:

\bea
\V - \Vp &=& \dV .
\lll{main}
\eea

\noi
This restriction is in terms of the reduced interaction and the change in the reduced interaction.

$\dV$ is defined in \reff{RFI}, which makes it clear that since $\Vp$ can be
expanded in powers of $\glap$, so can $\dV$:

\bea
\dV = \sum_{t=2}^\infty \glap^t \dVo{t} .
\lll{RFI exp}
\eea

\noi
We refer to $\dVo{t}$ as the $\OR(\glap^t)$ change in the reduced interaction, although for convenience 
the coupling
is factored out. Note that $\dVo{t}$ is a function of $\la$ and $\la'$.

Now \reff{main} can be expanded in powers of $\gla$ and $\glap$:

\bea
\sum_{t=1}^\infty \gla^t V^{(t)}(\la) - 
\sum_{t=1}^\infty \glap^t V^{(t)}(\la') = \sum_{t=2}^\infty \glap^t 
\dVo{t} .
\lll{u expand}
\eea

\noi
This equation is a bit tricky to use because it involves the coupling at two different scales.
To see how they are related, consider the matrix element of \reff{main} for $\phi_1 \rightarrow \phi_2 \phi_3$:

\bea
&&\left<\phi_2 \phi_3 \right| \V \left| \phi_1 \right> - \left<\phi_2 \phi_3 \right| \Vp\left| \phi_1 \right>  
= \left<\phi_2 \phi_3 \right| \dV \left| \phi_1 \right> .
\lll{gen 3p me}
\eea

\noi
According to the definition of the coupling, this equation implies 

\bea
&&\gla - \glap  = \left[ \SP p_1^+ \delta^{(5)}(p_1 - p_2 - p_3) \right]^{-1} \left<\phi_2 \phi_3 \right| \dV
\left| \phi_1 \right>\:\rule[-2mm]{.1mm}{6mm}_{\; \vec p_{2 \perp}
= \vec p_{3 \perp} = 0 \, ; \, 
p_2^+ = p_3^+ = \frac{1}{2} p_1^+} .
\lll{running 1}
\eea

\noi
Since $V^{(1)}$ changes particle number by 1, inspection of \reff{RFI} reveals that 
$\left<\phi_2 \phi_3 \right| \dV \left| \phi_1 \right>$ is $\OR(\glap^3)$; so

\bea
\gla = \glap + \OR(\glap^3) .
\eea

\noi
This implies

\bea
\gla = \glap + \sum_{s=3}^\infty \glap^s C_s(\la,\la') ,
\lll{scale dep}
\eea

\noi
where the $C_s$'s are functions of $\la$ and $\la'$.  In Section 7 we calculate $C_3$ explicitly.
For an integer $t \ge 1$, \reff{scale dep} implies

\bea
\gla^t = \glap^t + \sum_{s=2}^\infty \glap^{t+s} B_{t,s}(\la,\la') ,
\lll{scale dep 2}
\eea

\noi
where the $B_{t,s}$'s are also functions of $\la$ and $\la'$, and can be calculated in terms of the $C_s$'s by
raising \reff{scale dep} to the $t^{\hspace{.1mm}\mathrm{th}}$ power.  

We substitute \reff{scale dep 2} into \reff{u expand} and demand that it hold order-by-order in $\glap$.
At $\OR(\glap^r)$ ($r \ge 1$), this implies

\bea
&&V^{(r)}(\la) - 
V^{(r)}(\la') = \delta V^{(r)} - \sum_{s=2}^{r-1}  B_{r-s,s} V^{(r-s)}(\la),
\lll{coupled}
\eea

\noi
where $\delta V^{(1)} = 0$, and we define any sum to be zero if its upper limit is less than its lower limit.

We have restricted the Hamiltonian to respect approximate transverse locality,
which means that its matrix elements can be expanded in powers of transverse momenta.  This means that 
the matrix elements of $V^{(r)}(\la)$ can also be expanded in powers of transverse momenta.  
Each term in a transverse-momentum expansion of a matrix element $\left<F \right|
V^{(r)}(\la) \left| I \right>$ is either cutoff-dependent or cutoff-independent.  We define
$V^{(r)}_{\CD}(\la)$ to be the cutoff-dependent part of $V^{(r)}(\la)$, i.e. the part that 
produces the cutoff-dependent terms in transverse-momentum
expansions of matrix elements of $V^{(r)}(\la)$.  We define
$V^{(r)}_{\CI}$ to be the cutoff-independent part of $V^{(r)}(\la)$, i.e. the part 
that produces the cutoff-independent terms in transverse-momentum
expansions of matrix elements of $V^{(r)}(\la)$.  Then

\bea
V^{(r)}(\la) = V^{(r)}_{\CD}(\la) + V^{(r)}_{\CI} .
\eea

\noi
When we substitute this equation into \reff{coupled}, the cutoff-independent parts of the terms
on the left-hand side (LHS) cancel, leaving

\bea
&&V^{(r)}_{\CD}(\la) - 
V^{(r)}_{\CD}(\la') = \delta V^{(r)} - \sum_{s=2}^{r-1}  B_{r-s,s}
V^{(r-s)}(\la) .
\lll{CD coupled}
\eea

\noi
This equation can be used to calculate the $\OR(\gla^r)$ reduced interaction 
in terms of the lower-order reduced interactions.  Note that for $r=1$, the
RHS of the equation is zero, implying that the cutoff-dependent part of $V^{(1)}$ is zero.  
Inspection of \reff{O1} shows this to be the case.

In the remainder of this section, we summarize the results of Appendix D, which contains more details and rigor
than are necessary here.  

Recall that momentum conservation implies that any matrix element 
$\left<F \right| V(\la) \left| I \right>$ can be written as an expansion in unique products of
momentum-conserving delta functions.  This means that an arbitrary matrix element of 
\reff{CD coupled} can be expanded in products of
delta functions and thus is equivalent to a set of equations, one for each possible product of delta
functions.  Given approximate transverse locality, each of the resulting equations can be expanded
in powers of transverse momenta.  Matching the coefficients of powers of transverse momenta on either
side of these equations allows us to rigorously derive the following results (see Appendix D for details).

First, the cutoff-dependent part of the 
$\OR(\gla^r)$ reduced interaction is given in terms of lower-order reduced interactions by

\bea
\left<F \right| V^{(r)}_{\CD}(\la) \left| I \right> = \left[ \dVome{r} - \sum_{s=2}^{r-1}  B_{r-s,s}
\left<F \right| V^{(r-s)}(\la) \left| I \right>\right]_\lz ,
\lll{cc soln}
\eea

\noi
where ``$\la \; \mathrm{terms}$" means that the RHS is to be 
expanded in powers of transverse momenta and only the terms in the expansion that depend on $\la$ 
are to be kept.  Recall that $\dV$ is defined in \reff{RFI}.

Second, the cutoff-independent part of $V^{(r)}(\la)$ is the part with three interacting
particles and no transverse momentum dependence.  If there is no such part, then $V^{(r)}(\la)$
is completely determined by \reff{cc soln}.  Third, $V^{(r)}(\la)$ can have a cutoff-independent part
only if $r$ is odd.  Fourth, the coupling runs at odd orders; i.e. $C_s$ is zero if $s$ is even
[see \reff{scale dep}].  Fifth, there is no wave function renormalization at any order in perturbation theory
in our approach because this would violate the restrictions that we have placed on the Hamiltonian.  

The sixth and final result from Appendix D is that the cutoff-independent part of the 
$\OR(\gla^r)$ reduced interaction for odd $r \ge 3$ 
is given in terms of the cutoff-dependent part and lower-order reduced interactions by the integral 
equation\footnote{It is very difficult to prove that integral equations of this type have a unique
solution; so we simply assume it is true in this case.}

\bea
\left<F \right| V^{(r)}_{\CI} \left| I \right> &=& \frac{1}{B_{r,2}} \left[ \dVome{r+2} 
\:\rule[-4mm]{.1mm}{8mm}_{\; \mathrm{Ext. \;} \vec k_\perp \rightarrow 0} - \sum_{s=3}^{r+1}  B_{r+2-s,s}
\left<F \right| V^{(r+2-s)}_{\CI} \left| I \right> \right]_{\mathrm{3 \; int. \; 
part.}} \hspace{-.5in} .
\lll{CI cc soln}
\eea

\noi
``Ext. $\vec k_\perp \rightarrow 0$" means the limit in which the transverse momenta
in the external states are taken to zero.  ``$\mathrm{3 \; int. \; part.}$" means the 
products of momentum-conserving delta functions that appear on the
RHS are to be examined and only the contributions from three interacting particles 
are to be kept.  

\reff{CI cc soln} is an integral equation because $V_{\CI}^{(r)}$ is nested with lower-order reduced interactions in
integrals
in $\delta V^{(r+2)}$.  A cursory examination of the definition of $\delta V^{(r+2)}$ may lead  one to believe
that
\reff{CI cc soln} is useless because it requires one to know $V^{(r+1)}(\la')$.  However,  since $r$ is odd,
$V^{(r+1)}(\la')=V^{(r+1)}_{\CD}(\la')$, and the dependence of $\delta V^{(r+2)}$ on $V^{(r+1)}_{\CD}(\la')$  can be replaced with
further
dependence on $V^{(r)}_{\CI}$ and $V^{(r)}_{\CD}(\la')$ (and lower-order reduced interactions) 
using \reff{cc soln} with $r \rightarrow r+1$.

In the remaining sections, we apply the results of this section to calculate a number of example matrix elements
of the Hamiltonian, demonstrating that they are indeed uniquely determined in terms of the fundamental parameters
of the theory by the restrictions that the
Hamiltonian must produce cutoff-independent physical quantities and be consistent with the unviolated physical
principles of the theory.

\section{Second-Order Matrix Elements}

In this section we illustrate our method by calculating some second-order matrix elements of the Hamiltonian.
The matrix elements of the Hamiltonian are defined in terms of the matrix elements of the reduced interaction by
\reff{cutoff}.   To calculate matrix elements of the Hamiltonian to second-order, we need to
calculate the corresponding matrix elements of $V^{(2)}(\la)$, the second-order reduced interaction.  As mentioned in
the previous section, for $V^{(r)}(\la)$
to have a cutoff-independent part, $r$ must be odd. According to \reff{cc soln}, this means that $V^{(2)}(\la)$
is given in terms of the second-order change to the reduced interaction by

\bea
\left<F \right| V^{(2)}(\la) \left| I \right> &=& \dVome{2}\lzb ,
\lll{omega2 from R}
\eea

\noi
where the second-order change to the reduced interaction is defined in terms of the first-order
reduced interaction and the cutoff function:

\bea
\dVome{2} &=& \frac{1}{2} \sum_K
\left<F \right| V^{(1)} \left| K \right> \left<K \right| V^{(1)} \left| I \right> T_2^{(\la,\la')}(F,K,I) .
\lll{gen 2}
\eea

\subsection{Example: $\left< \phi_2 \right| \M \left| \phi_1 \right>$}

As a first example, we compute the matrix element $\left< \phi_2 \right| \M \left| \phi_1 \right>$ to
$\OR(\gla^2)$.  To compute this matrix element, we use $\left| I \right> = \left| \phi_1 \right>$ and
$\left| F \right> = \left| \phi_2 \right>$.  Then the matrix element of the 
second-order change to the reduced interaction is

\bea
\left< \phi_2 \right| \delta V^{(2)} \left| \phi_1 \right>
&=& \frac{1}{2} \sum_K
\left<\phi_2 \right| V^{(1)} \left| K \right> \left<K \right| V^{(1)} \left| \phi_1 \right>
T_2^{(\la,\la')}(F,K,I) .
\lll{one body}
\eea

\noi
Since $V^{(1)}$ changes particle number by 1, $\left| K \right>$ has to be a two-particle state 
and \reff{one body} becomes

\bea
\left< \phi_2 \right| \delta V^{(2)} \left| \phi_1 \right>  &=& \frac{1}{4} \int D_3 D_4
\left<\phi_2 \right| V^{(1)} \left| \phi_3 \phi_4 \right> \left<\phi_3 \phi_4 \right| V^{(1)} \left| \phi_1
\right> 
T_2^{(\la,\la')}(F,K,I),
\lll{2p me loop 1}
\eea

\noi
where the completeness relation in \reff{completeness} is used (see Appendix B for conventions and definitions)
and $\left| K \right> = \left| \phi_3 \phi_4
\right>$.  Here $\left< \phi_2 \right| \delta V^{(2)} \left| \phi_1 \right>$ is represented diagrammatically in Fig.
1.  
\begin{figure}
\centerline{\epsffile{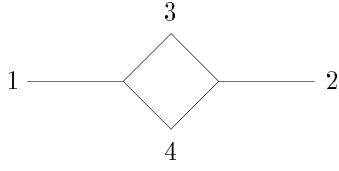}}
\caption{A diagrammatic representation of $\left< \phi_2 \right| \delta V^{(2)} \left| \phi_1
\right>$, a matrix element of the second-order change to the reduced 
interaction. The numbers label the particles.}
\end{figure}
Using the expression for $V^{(1)}$ in \reff{O1},

\bea
\left< \phi_2 \right| \delta V^{(2)} \left| \phi_1 \right>  &=& \frac{1}{4} 
\int D_3 D_4 \D{2}{3 4} \D{1}{3 4} T_2^{(\la,\la')}(F,K,I) .
\lll{2p me loop 2}
\eea

\noi
The integral over $p_3$ can be done with one of the delta functions.  At this point it is convenient to change 
variables from $p_4^+$ and $\vec p_{4 \perp}$ to $x$ and $\vec r_\perp$:

\bea
p_4 = (p_4^+,\vec p_{4 \perp}) = (x p_1^+, x \vec p_{1 \perp} + \vec r_\perp) ,
\eea

\noi
where $x$ is the fraction of the total longitudinal momentum carried by particle 4 and
$\vec r_\perp$ is the transverse momentum of particle 4 in the center-of-mass frame.  
Since $p_4^-$ is constrained, we do not display it in the list of
components of $p_4$.  The momentum-conserving delta functions imply

\bea
p_3 = \left([1-x] p_1^+, [1-x] \vec p_{1 \perp} - \vec r_\perp \right) .
\eea

\noi
Note that all longitudinal momenta are positive, although we do not explicitly show these limits in integrals.
This implies, for example, that $0 < x < 1$.  With the change of variables, \reff{2p me loop 2} becomes

\bea
\left< \phi_2 \right| \delta V^{(2)} \left| \phi_1 \right>  &=& \frac{1}{4}  p_1^+ \delta^{(5)}(p_2 - p_1)\int 
d^4 r_\perp dx \frac{1}{x (1-x)} T_2^{(\la,\la')}(F,K,I) .
\lll{2p me loop 3}
\eea

We are using massless particles; so the free masses of the states $\left| I \right> = \left| \phi_1 \right>$ and
$\left| F \right> = \left| \phi_2 \right>$ are zero:

\bea
M^2_I = M^2_F = 0.
\eea

\noi
The particles in the intermediate state $\left| K \right> = \left| \phi_3 \phi_4 \right>$ 
are also massless, but the state still has a free mass from the particles' relative motion:

\bea
M^2_K = \frac{\rsq}{x (1-x)} .
\eea

\noi
Then from the definition of the cutoff function $T_2^{(\la,\la')}$,

\bea
T_2^{(\la,\la')}(F,K,I) = - \frac{2 x (1-x)}{\rsq} \left( e^{-2 \la'^{-4} \frac{\vec r^{\, 4}_\perp}{x^2
(1-x)^2}} - 
e^{-2 \la^{-4} \frac{\vec r^{\, 4}_\perp}{x^2 (1-x)^2}} \right) .
\eea

\noi
After the integral is done, \reff{2p me loop 3} becomes

\bea
\left< \phi_2 \right| \delta V^{(2)} \left| \phi_1 \right>  &=& \D{1}{2}
\frac{1}{1536 \pi^3}
\sqrt{\frac{\pi}{2}} \left( \la^2 - \la'^2 \right).
\lll{2p me loop 4}
\eea

\noi
Applying \reff{omega2 from R} yields the
matrix element of the second-order reduced interaction:

\bea
\left<\phi_2 \right| V^{(2)}(\la) \left| \phi_1 \right>  &=& \D{1}{2} \la^2
\frac{1}{1536 \pi^3}
\sqrt{\frac{\pi}{2}} .
\lll{self energy 1}
\eea

\noi
Using the definition of the Hamiltonian in terms of the reduced interaction,
we find that to second-order, the matrix element of the Hamiltonian is given by

\bea
\left<\phi_2 \right| \M \left| \phi_1 \right> = \D{1}{2} \la^2 \;
\gla^2 \frac{1}{1536 \pi^3} \sqrt{\frac{\pi}{2}} .
\lll{one body result}
\eea

This matrix element of the Hamiltonian has been completely determined to second-order, 
in terms of the fundamental parameters of the theory, by the restrictions that the
Hamiltonian has to produce cutoff-independent physical quantities and has to respect the unviolated physical
principles of the
theory.
If we wanted to compute physical quantities, this is one of the matrix elements of the Hamiltonian that we might
need.  We would
have to choose $\la$ and 
determine $\gla$ by fitting data since the theory contains one adjustable parameter.
Note that \reff{one body result} is consistent with all the restrictions that we have placed on the Hamiltonian.

\subsection{Example: $\left<\phi_3 \phi_4 \right| \M \left| \phi_1 \phi_2 \right>$}

Next we calculate the matrix element 
$\left<\phi_3 \phi_4 \right| \M \left| \phi_1 \phi_2 \right>$ to $\OR(\gla^2)$.  We use 
$\left| I \right> = \left| \phi_1 \phi_2 \right>$ and
$\left| F \right> = \left| \phi_3 \phi_4 \right>$.  Then the matrix element of the second-order change
to the reduced interaction is

\bea
\dVotwototwo{2}
&=& \frac{1}{2} \sum_K
\left<\phi_3 \phi_4 \right| V^{(1)} \left| K \right> \left<K \right| V^{(1)} \left| \phi_1 \phi_2  \right> 
T_2^{(\la,\la')}(F,K,I) .
\lll{two body}
\eea

\noi
Since $V^{(1)}$ changes particle number by 1, $\left| K \right>$ has to be either a one- or 
three-particle state.  When $V^{(1)}$ is
substituted into the RHS of this equation and all the creation and annihilation operators from the possible
intermediate states and the interactions are contracted, we see that $\dVotwototwo{2}$ 
has contributions from a number of different processes. We define
$\dVotwototwo{2}_i$ to be the contribution to $\dVotwototwo{2}$ from the process shown in Fig. 2-i, 
where $1 \leq i \leq 9$.  
\begin{figure}
\centerline{\epsffile{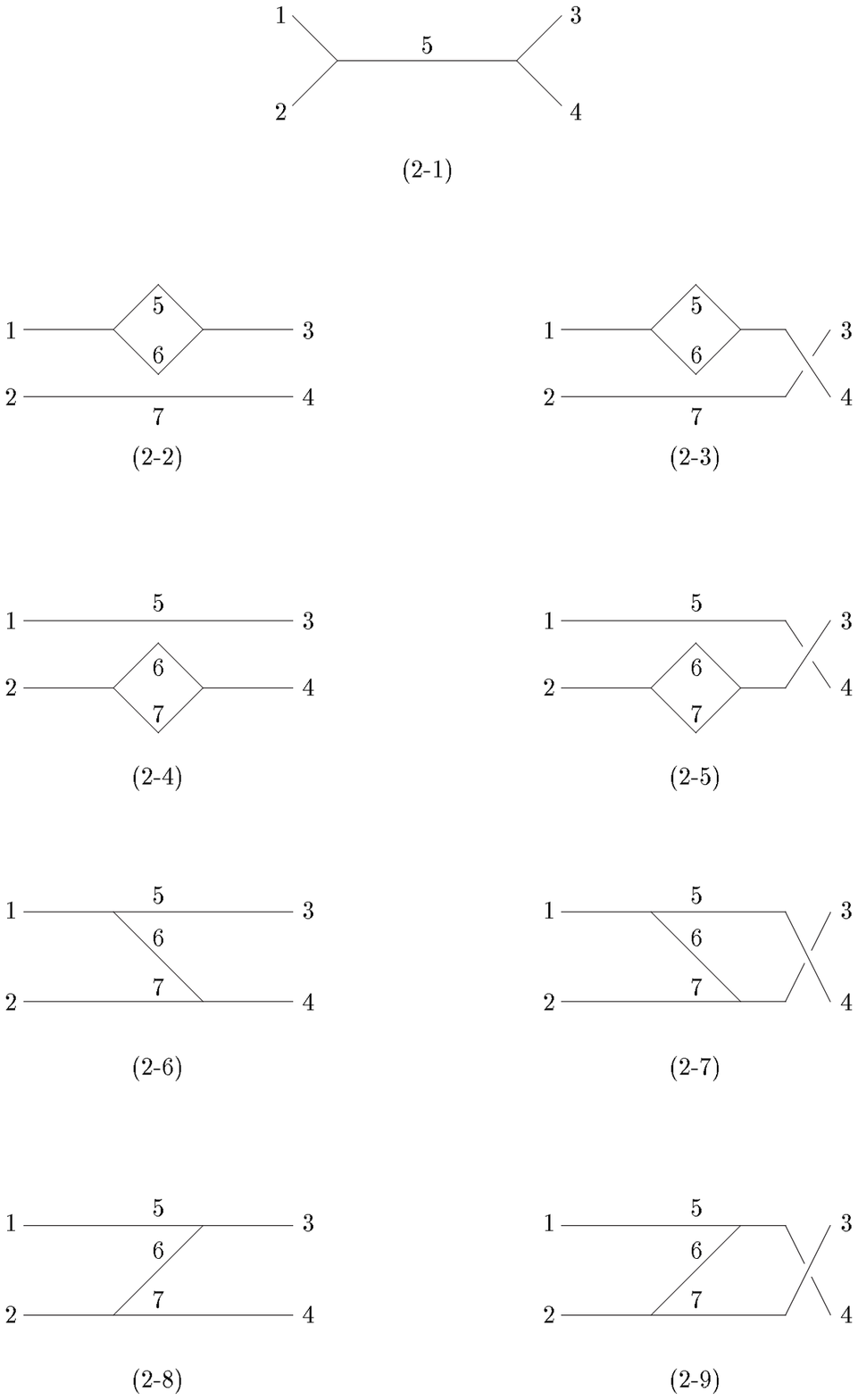}}
\caption{A diagrammatic representation of the contributions to $\dVotwototwo{2}$, a matrix element of the
second-order change to the reduced interaction.  The numbers label the particles.}
\end{figure}
Then

\bea
\dVotwototwo{2} = \sum_{i=1}^9 \dVotwototwo{2}_i .
\lll{R sum}
\eea

\reff{omega2 from R} suggests that we define a contribution to the matrix element of the second-order reduced
interaction
for each contribution to $\dVotwototwo{2}$:

\bea
\left<\phi_3 \phi_4 \right| V^{(2)}(\la) \left| \phi_1 \phi_2 \right>_i &=& \dVotwototwo{2}_i \lzb ,
\lll{me part def}
\eea

\noi
and then

\bea
\left<\phi_3 \phi_4 \right| V^{(2)}(\la) \left| \phi_1 \phi_2 \right>
&=& \sum_{i=1}^9 \left<\phi_3 \phi_4 \right| V^{(2)}(\la) \left| \phi_1 \phi_2 \right>_i .
\lll{two body break}
\eea

For each contribution to $\dVotwototwo{2}$, we use the change of variables

\bea
p_1 &=& (x P^+, x \vec P_\perp + \vec r_\perp) , \n
p_2 &=& ([1-x] P^+, [1-x] \vec P_\perp - \vec r_\perp) , \n
p_3 &=& (y P^+, y \vec P_\perp + \vec q_\perp) , \n
p_4 &=& ([1-y] P^+, [1-y] \vec P_\perp - \vec q_\perp) ,
\eea

\noi
where $P=p_1+p_2$.  Then the initial and final free masses are

\bea
M_I^2 &=& \frac{\rsq}{x (1-x)} ,\n
M_F^2 &=& \frac{\qsq}{y (1-y)} .\n
\eea

\subsubsection{The Annihilation Contribution}

The contribution to $\dVotwototwo{2}$ from the annihilation process, shown in Fig. 2-1, is

\bea
\dVotwototwo{2}_1  &=& \frac{1}{2} \int D_5
\left<\phi_3 \phi_4 \right| V^{(1)} \left| \phi_5 \right> \left<\phi_5 \right| V^{(1)} \left| \phi_1 \phi_2
\right> 
T_2^{(\la,\la')}(F,K,I) \n
&=& \D{1 2}{3 4} \frac{1}{2} 
\left(\frac{1}{M_F^2} + \frac{1}{M_I^2} \right) 
\left( e^{-2 \la'^{-4} M_F^2 M_I^2} - e^{-2 \la^{-4} M_F^2 M_I^2} 
\right) ,
\eea

\noi
where the intermediate state is $\left| K \right> = \left| \phi_5 \right>$.  According to \reff{me part def}, the corresponding part
of the matrix element of the second-order reduced interaction is

\bea
\left<\phi_3 \phi_4 \right| V^{(2)}(\la) \left| \phi_1 \phi_2 \right>_1  &=& \D{1 2}{3 4} \frac{1}{2} 
\left(\frac{1}{M_F^2} + \frac{1}{M_I^2} \right) 
\left( e^{-2 \la'^{-4} M_F^2 M_I^2} - e^{-2 \la^{-4} M_F^2 M_I^2} 
\right) \lzb \hspace{-.3in}.
\lll{ann 1}
\eea

If we expand everything multiplying the delta function in powers of transverse momenta, the lowest order terms
from the two exponentials cancel
and leave two types of terms: those depending on $\la$ and those depending on $\la'$.  We can remove the
$\la'$-dependent terms without altering the cancellation of the lowest term 
by replacing the first exponential with a $1$:

\bea
\left<\phi_3 \phi_4 \right| V^{(2)}(\la) \left| \phi_1 \phi_2 \right>_1  &=& \D{1 2}{3 4} \frac{1}{2}
\left(\frac{1}{M_F^2} + \frac{1}{M_I^2} \right) 
\left( 1 - e^{-2 \la^{-4} M_F^2 M_I^2} 
\right) .
\lll{ann result}
\eea

\subsubsection{The Two-Particle Self-Energy Contributions}

The rest of the contributions to $\dVotwototwo{2}$ have a three-body intermediate state,
$\left| K \right> = \left| \phi_5 \phi_6 \phi_7 \right>$.  The contribution from the two-particle self-energy,
shown in Fig. 2-2, is given by

\bea
\dVotwototwo{2}_2  &=& \frac{1}{4} \frac{1}{x y}
\int D_5 \, D_6 \, D_7 \, T_2^{(\la,\la')}(F,K,I) \D{3}{5 6} \D{7}{4} \D{1}{6 5} \D{7}{2}  .
\eea

\noi
We use the change of variables

\bea
p_5 &=& (z p_1^+, z \vec p_{1 \perp} + \vec k_\perp) ,
\eea

\noi
and do the integrals over $p_6$ and $p_7$ using the delta functions.  Then 
we find that

\bea
p_6 &=& ([1-z] p_1^+, [1-z] \vec p_{1 \perp} - \vec k_\perp) ,
\eea

\noi
and the free mass of the intermediate state is

\bea
M_K^2 &=& \frac{\rsq}{x(1-x)} + \frac{\ksq}{z (1-z) x} ,
\eea

\noi
and

\bea
\dVotwototwo{2}_2  &=& \frac{1}{2} \frac{1}{\SP} \D{1}{3} 
\D{2}{4} \frac{1}{x} \int d^4 k_\perp dz \frac{1}{\ksq} \left[ e^{-2 \la^{-4} \bx{2}{IK}} - e^{-2 \la'^{-4}
\bx{2}{IK}} 
\right] \n
&=& \D{1}{3} \D{2}{4} \frac{1}{1536 \pi^3} \sqrt{\frac{\pi}{2}} \left( \la^2 - \la'^2 \right).
\eea

\noi
According to \reff{me part def}, the corresponding part of the matrix element of the 
second-order reduced interaction is 

\bea
\left<\phi_3 \phi_4 \right| V^{(2)}(\la) \left| \phi_1 \phi_2 \right>_2  &=& 
\D{1}{3} \D{2}{4} \la^2 \frac{1}{1536 \pi^3} \sqrt{\frac{\pi}{2}} .
\lll{two self}
\eea

\noi
Note that Eqs. (\ref{eq:self energy 1}) and (\ref{eq:two self}) indicate that to $\OR(\gla^2)$ in this theory the
only 
effect a spectator has on the self-energy is to produce an extra delta function.  This is what one
would expect if cluster decomposition were maintained.  However, there is no guarantee that this
will hold to all orders, and it is known that in gauge theories
$\left<\phi_3 \phi_4 \right| V^{(2)}(\la) \left| \phi_1 \phi_2 \right>_2$
depends on $p_2^+/P^+$.  This is allowed because our cutoff partially violates
cluster decomposition.

According to Fig. 2,

\bea
\left<\phi_3 \phi_4 \right| V^{(2)}(\la) \left| \phi_1 \phi_2 \right>_3 &=& \left<\phi_3 \phi_4 \right|
V^{(2)}(\la) \left| \phi_1 \phi_2
\right>_2 \:\rule[-4mm]{.1mm}{8mm}_{\; 3 \leftrightarrow 4} = \D{1}{4} \D{2}{3} \la^2 \frac{1}{1536
\pi^3} \sqrt{\frac{\pi}{2}} , \n\n
\left<\phi_3 \phi_4 \right| V^{(2)}(\la) \left| \phi_1 \phi_2 \right>_4 &=& \left<\phi_3 \phi_4 \right|
V^{(2)}(\la) \left| \phi_1 \phi_2
\right>_2 \:\rule[-4mm]{.1mm}{8mm}_{\; 3 \leftrightarrow 4}^{\; 1 \leftrightarrow 2} = \D{1}{3} \D{2}{4} 
\la^2 \frac{1}{1536 \pi^3} \sqrt{\frac{\pi}{2}} , \n\n
\left<\phi_3 \phi_4 \right| V^{(2)}(\la) \left| \phi_1 \phi_2 \right>_5 &=& \left<\phi_3 \phi_4 \right|
V^{(2)}(\la) \left| \phi_1 \phi_2
\right>_4 \:\rule[-4mm]{.1mm}{8mm}_{\; 3 \leftrightarrow 4} = \D{1}{4} \D{2}{3} \la^2 \frac{1}{1536
\pi^3} \sqrt{\frac{\pi}{2}} .
\lll{4 pt se}
\eea

\subsubsection{The Exchange Contributions}

The exchange contribution to the matrix element of the 
second-order change in the reduced interaction, shown in Fig. 2-6, is

\bea
\dVotwototwo{2}_6 &=& \frac{1}{2} \frac{1}{x (1-y)}\int D_5
D_6 
D_7 T_2^{(\la,\la')}(F,K,I) \D{4}{6 7} \D{7}{2} \D{1}{5 6} \D{5}{3} \n
&=& - \D{1 2}{3 4} \frac{1}{2} \frac{1}{x-y}
\left[ \frac{1}{\Bx{}{KF}} + \frac{1}{\Bx{}{KI}} \right] 
\left[ e^{-2 \la'^{-4} \bx{}{KF} \bx{}{KI} } - e^{-2 \la^{-4} \bx{}{KF} \bx{}{KI} } \right]
,
\eea

\noi
where the free mass of the intermediate state is

\bea
M_K^2 &=& \frac{\rsq}{1-x} + \frac{\qsq}{y} + \frac{(\vec r_\perp - \vec q_\perp)^2}{x - y} .
\eea

\noi
\reff{me part def} implies that the corresponding part of the matrix element of the 
second-order reduced interaction is given by

\bea
&&\left<\phi_3 \phi_4 \right| V^{(2)}(\la) \left| \phi_1 \phi_2 \right>_6 = - 
\D{1 2}{3 4} \frac{1}{2} \frac{1}{x-y}
\left[ \frac{1}{\Bx{}{KF}} + \frac{1}{\Bx{}{KI}} \right] 
\left[ 1 - e^{-2 \la^{-4} \bx{}{KF} \bx{}{KI} } \right] .
\lll{exchange}
\eea

\noi
According to Fig. 2, the remaining parts of the matrix element of the 
second-order reduced interaction are given by

\bea
\left<\phi_3 \phi_4 \right| V^{(2)}(\la) \left| \phi_1 \phi_2 \right>_7 &=& \left<\phi_3 \phi_4 \right|
V^{(2)}(\la) \left| \phi_1 \phi_2
\right>_6 \:\rule[-4mm]{.1mm}{8mm}_{\; 3 \leftrightarrow 4} = \left<\phi_3 \phi_4 \right| V^{(2)}(\la) \left|
\phi_1 \phi_2
\right>_6 \:\rule[-4mm]{.1mm}{8mm}_{\; y \rightarrow 1-y \, , \, \vec q_\perp \rightarrow - \vec q_\perp} , \n\n
\left<\phi_3 \phi_4 \right| V^{(2)}(\la) \left| \phi_1 \phi_2 \right>_8 &=& \left<\phi_3 \phi_4 \right|
V^{(2)}(\la) \left| \phi_1 \phi_2
\right>_6 \:\rule[-4mm]{.1mm}{8mm}_{\; 3 \leftrightarrow 4}^{\; 1 \leftrightarrow 2} = 
\left<\phi_3 \phi_4 \right| V^{(2)}(\la) \left| \phi_1 \phi_2
\right>_6 \:\rule[-4mm]{.1mm}{8mm}_{\; y \rightarrow 1-y \, , \, 
\vec q_\perp \rightarrow - \vec q_\perp}^{\; x \rightarrow 1-x \, , \, \vec r_\perp \rightarrow - \vec r_\perp} ,
\n\n
\left<\phi_3 \phi_4 \right| V^{(2)}(\la) \left| \phi_1 \phi_2 \right>_9 &=& \left<\phi_3 \phi_4 \right|
V^{(2)}(\la) \left| \phi_1 \phi_2
\right>_8 \:\rule[-4mm]{.1mm}{8mm}_{\; 3 \leftrightarrow 4} = \left<\phi_3 \phi_4 \right| V^{(2)}(\la) \left|
\phi_1 \phi_2
\right>_6 \:\rule[-4mm]{.1mm}{8mm}_{\; x \rightarrow 1-x \, , \, \vec r_\perp \rightarrow - \vec r_\perp}
\hspace{-.5in}.
\lll{other exchange}
\eea

\subsubsection{The Complete Matrix Element and Transverse Locality}

Using Eqs. (\ref{eq:cutoff}) and (\ref{eq:order exp}), 
we can write the full second-order matrix element of the Hamiltonian:

\bea
\left<\phi_3 \phi_4 \right| \M \left| \phi_1 \phi_2 \right> &=& (\D{1}{3} \D{2}{4} + \D{1}{4} \D{2}{3} )
M_F^2 + e^{- \la^{-4} \bx{2}{FI}} \gla^2 \sum_{i=1}^9 \left<\phi_3 \phi_4 \right| V^{(2)}(\la) \left| \phi_1
\phi_2 \right>_i .
\lll{2 to 2 result}
\eea

\noi
By inspection, it is clear that this result respects all the restrictions we
have placed on the Hamiltonian, except perhaps the requirement that 
it can be represented as a power series in transverse momenta with an infinite
radius of convergence.  To see how to check this, consider the first term in the sum on the RHS:

\bea
e^{- \la^{-4} \bx{2}{FI}} \gla^2 \left<\phi_3 \phi_4 \right| V^{(2)}(\la) \left| \phi_1
\phi_2 \right>_1 &=& \D{1 2}{3 4} \frac{1}{2} \gla^2 e^{- \la^{-4} \left( M_F^2 - M_I^2 \right)^2}  
\left(\frac{1}{M_F^2} + \frac{1}{M_I^2} \right) \n
&\times& \left( 1 - e^{-2 \la^{-4} M_F^2 M_I^2} \right) .
\lll{local test 1}
\eea

\noi
The presence of the mass denominators might seem likely to cause a problem with the convergence of a power
series expansion.
Everything multiplying the delta function can be expressed as a power series in transverse momenta if
it can be expressed as a power series in $M_F^2$ and $M_I^2$, since it is a function
only of these and since $M_F^2$ and $M_I^2$ are themselves power series in transverse momenta.
We can write

\bea
&&e^{- \la^{-4} \bx{2}{FI}} \gla^2 \left<\phi_3 \phi_4 \right| V^{(2)}(\la) \left| \phi_1
\phi_2 \right>_1 = \D{1 2}{3 4} \frac{1}{2} \gla^2 \frac{1}{\la^2} \bigg[ \frac{2}{\la^2} (M_F^2 + M_I^2)
- \frac{2}{\la^6} (M_F^6 + M_I^6) \n
&+& \frac{1}{\la^{10}} \left\{ \frac{4}{3} (M_F^4 M_I^6 + M_F^6 M_I^4)
- (M_F^8 M_I^2 + M_F^2 M_I^8) \right\} + \cdots \bigg] .
\lll{local test 2}
\eea

\noi
If we approximate the RHS of this equation by keeping only $N$ terms, the difference between
this finite sum and the RHS of \reff{local test 1} can be made arbitrarily small, for any values of
$M_F^2$ and $M_I^2$, by making $N$ large enough.  This means that $e^{- \la^{-4} \bx{2}{FI}} 
\gla^2 \left<\phi_3 \phi_4 \right| V^{(2)}(\la) \left| \phi_1
\phi_2 \right>_1$ can indeed be represented as a power series in transverse momenta with an infinite
radius of convergence.  The other contributions to $\left<\phi_3 \phi_4 \right|
\M \left| \phi_1 \phi_2 \right>$ can be analyzed similarly, and the conclusion is the same for each.

\subsection{Example: $\left< \phi_2 \phi_3 \phi_4 \right| \M \left| \phi_1 \right>$}

The final second-order matrix element of the Hamiltonian that we calculate is 
$\left< \phi_2 \phi_3 \phi_4 \right| \M \left| \phi_1 \right>$.  To compute this matrix element, we
use $\left| I \right> = \left| \phi_1 \right>$ and $\left| F \right> = \left| \phi_2 \phi_3 \phi_4 \right>$.
The matrix element of the second-order change to the reduced interaction is

\bea
\dVoonetothree{2}
&=& \frac{1}{2} \sum_K
\left<\phi_2 \phi_3 \phi_4 \right| V^{(1)} \left| K \right> \left<K \right| V^{(1)} \left| \phi_1 \right> 
T_2^{(\la,\la')}(F,K,I) .
\lll{1 to 3}
\eea

\noi
Since $V^{(1)}$ changes particle number by 1,  
$\left| K \right>$ has to be a two-particle state.  

We define 
$\dVoonetothree{2}_i$ to be the contribution to $\dVoonetothree{2}$ from the process shown in 
Fig. 3-i, where $1 \leq i \leq 3$.  
\begin{figure}
\centerline{\epsffile{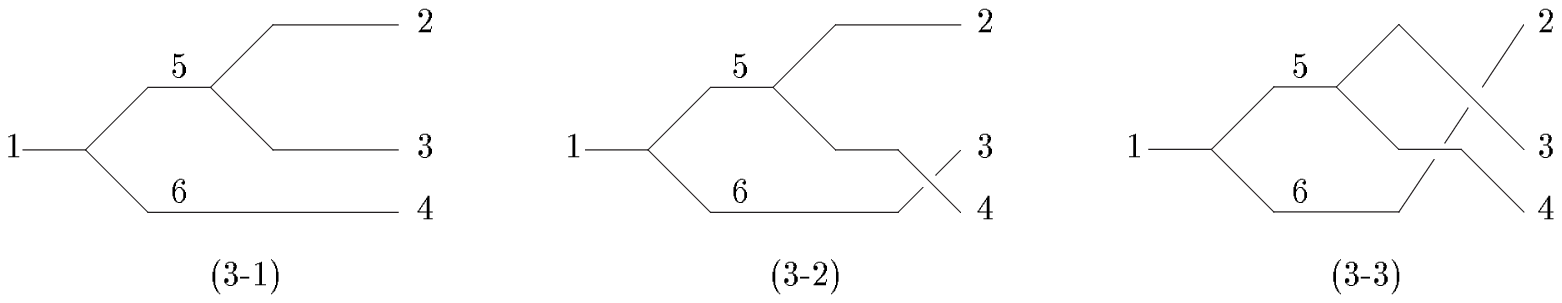}}
\caption{A diagrammatic representation of the contributions to 
$\dVoonetothree{2}$, a matrix element of the second-order change to the reduced interaction.
The numbers label the particles.}
\end{figure}
Then

\bea
\dVoonetothree{2} = \sum_{i=1}^3 \dVoonetothree{2}_i .
\lll{R sum2}
\eea

\noi
\reff{omega2 from R} suggests that we define a contribution to the matrix element of the second-order reduced
interaction
for each contribution to $\dVoonetothree{2}$:

\bea
\left<\phi_2 \phi_3 \phi_4 \right| V^{(2)}(\la) \left| \phi_1 \right>_i &=& \dVoonetothree{2}_i \lzb ,
\lll{me part def2}
\eea

\noi
and then

\bea
\left<\phi_2 \phi_3 \phi_4 \right| V^{(2)}(\la) \left| \phi_1 \right>
&=& \sum_{i=1}^3 \left<\phi_2 \phi_3 \phi_4 \right| V^{(2)}(\la) \left| \phi_1 \right>_i .
\eea

For each contribution to $\dVoonetothree{2}$, we use $\left| K \right> = \left| \phi_5 \phi_6 \right>$ 
and the change of variables

\bea
p_2 &=& (x p_1^+, x \vec p_{1 \perp} + \vec r_\perp) , \n
p_3 &=& (y p_1^+, y \vec p_{1 \perp} + \vec q_\perp) , \n
p_4 &=& (z p_1^+, z \vec p_{1 \perp} + \vec w_\perp) ,
\eea

\noi
where 

\bea
x+y+z &=& 1 ,\n
\vec r_\perp + \vec q_\perp + \vec w_\perp  &=& 0 .
\eea

\noi
Then the relevant differences of free masses are

\bea
\Bx{}{FK} &=& \frac{\rsq}{x} + \frac{\qsq}{y} - \frac{\wsq}{x+y} , \n
\Bx{}{KI} &=& \wsq \left( \frac{1}{x+y} + \frac{1}{z} \right) , \n
\Bx{}{FI} &=& \frac{\rsq}{x} + \frac{\qsq}{y} + \frac{\wsq}{z} , \n
\eea

\noi
and the contribution to the matrix element of the second-order change to the reduced interaction shown
in Fig. 3-1 is

\bea
\dVoonetothree{2}_1  &=& \D{1}{2 3 4} \frac{1}{2}
\frac{1}{x+y}
\left(\frac{1}{\Bx{}{FK}} - \frac{1}{\Bx{}{KI}} \right) 
\left( e^{2 \la'^{-4} \bx{}{FK} \bx{}{KI}} - e^{2 \la^{-4} \bx{}{FK} \bx{}{KI}} 
\right) ,
\eea

\noi
and the corresponding contribution to the matrix element of the second-order reduced interaction is

\bea
\left<\phi_2 \phi_3 \phi_4 \right| V^{(2)}(\la) \left| \phi_1 \right>_1  &=& \D{1}{2 3 4} \frac{1}{2}
\frac{1}{x+y}
\left(\frac{1}{\Bx{}{FK}} - \frac{1}{\Bx{}{KI}} \right) \n
&\times& 
\left( e^{2 \la'^{-4} \bx{}{FK} \bx{}{KI}} - e^{2 \la^{-4} \bx{}{FK} \bx{}{KI}} 
\right) \lzb .
\eea

As before, if we expand everything multiplying the delta function in powers of transverse momenta, 
the lowest order terms from the two exponentials cancel
and leave two types of terms: those depending on $\la$ and those depending on $\la'$.  We can remove the
$\la'$-dependent terms without altering the cancellation of the lowest term 
by replacing the first exponential with a $1$:

\bea
\left<\phi_2 \phi_3 \phi_4 \right| V^{(2)}(\la) \left| \phi_1 \right>_1  &=& \D{1}{2 3 4} \frac{1}{2}
\frac{1}{x+y}
\left(\frac{1}{\Bx{}{FK}} - \frac{1}{\Bx{}{KI}} \right) 
\left( 1 - e^{2 \la^{-4} \bx{}{FK} \bx{}{KI}} 
\right).
\eea

According to Fig. 3, the other parts of the matrix element of the second-order reduced interaction are given by

\bea
\left<\phi_2 \phi_3 \phi_4 \right| V^{(2)}(\la) \left| \phi_1 \right>_2 &=& \left<\phi_2 \phi_3 \phi_4 \right|
V^{(2)}(\la) \left| \phi_1
\right>_1\:\rule[-4mm]{.1mm}{8mm}_{\; 3 \leftrightarrow 4} = \left<\phi_2 \phi_3 \phi_4 \right| V^{(2)}(\la)
\left| \phi_1
\right>_1\:\rule[-4mm]{.1mm}{8mm}_{\; y \leftrightarrow z \; , \; \vec q_\perp \leftrightarrow \vec w_\perp}
\hspace{-.3in} ,
\eea

\noi
and

\bea
\left<\phi_2 \phi_3 \phi_4 \right| V^{(2)}(\la) \left| \phi_1 \right>_3 &=& \left<\phi_2 \phi_3 \phi_4 \right|
V^{(2)}(\la) \left| \phi_1
\right>_1\:\rule[-4mm]{.1mm}{8mm}_{\; 2 \rightarrow 3 \; , \; 3 \rightarrow 4 \; , \; 4 \rightarrow 2} \n
&=& \left<\phi_2 \phi_3 \phi_4 \right| V^{(2)}(\la) \left| \phi_1 \right>_1
\:\rule[-4mm]{.1mm}{8mm}^{\; x \rightarrow y \; , \; y \rightarrow z \; , \; z \rightarrow x}_{\; \vec r_\perp 
\rightarrow \vec q_\perp \; , \;
\vec q_\perp \rightarrow \vec w_\perp \; , \; \vec w_\perp \rightarrow \vec r_\perp } .
\eea

\noi
Using the definition of the Hamiltonian in terms of the reduced interaction, we can write
the full second-order matrix element of the Hamiltonian:

\bea
\left<\phi_2 \phi_3 \phi_4 \right| \M \left| \phi_1 \right> &=& e^{-\la^{-4} M_F^4} \gla^2 \sum_{i=1}^3 
\left<\phi_2 \phi_3 \phi_4 \right| V^{(2)}(\la) \left| \phi_1 \right>_i .
\eea

\noi
This result is consistent with the restrictions we have placed on the Hamiltonian.  

\section{The Removal of Cutoff Dependence from Physical Quantities}

One of the restrictions on the Hamiltonian is that it has to produce cutoff-independent physical quantities.
To see how this requirement is fulfilled, consider as an example 
the scattering cross section for $\phi_1 \phi_2 \rightarrow \phi_3 \phi_4$.  At second order,
the $T$ matrix has contributions from the $s$, $t$, and $u$ channels.  The different channels are
linearly independent  functions of the external momenta; so each contribution should individually have all the
properties required of the full $T$ matrix.  For simplicity, we limit ourselves to consideration of the $s$
(annihilation) channel.

In our approach, the scattering matrix $S$ is defined by

\bea
\left< F \right| S \left| I \right> = \left< F \right| \left. \! I \right> - 2 \pi i \delta(P_F^- - P_I^-)
\left< F \right| T(P_I^-) \left| I \right>,
\eea

\noi
where the $T$ matrix is defined by [16]

\bea
T(p^-) = H_I(\la) + H_I(\la) \frac{1}{p^- - h + i \epsilon} T(p^-) ,
\lll{T}
\eea

\noi
and $H_I(\la)$ is the interacting part of the Hamiltonian cut off by $\la$:

\bea
H_I(\la) = \frac{\MI}{{\cal P}^+} .
\lll{hamm}
\eea

\noi
To second order,

\bea
\left< \phi_3 \phi_4 \right| T(p_1^- + p_2^-) \left| \phi_1 \phi_2 \right> &=& 
\left< \phi_3 \phi_4 \right|  H_I(\la) + 
H_I(\la) \frac{1}{p_1^- + p_2^- - h + i \epsilon} H_I(\la) \left| \phi_1 \phi_2 \right> .
\lll{T exp}
\eea

\noi
If we neglect all noncanonical operators, the first term does not contribute and the second term can be
calculated by inserting a complete set of free states.  This leads to

\bea
\delta(P_F^- - P_I^-) \left< \phi_3 \phi_4 \right| T(P_I^-) \left| \phi_1 \phi_2 \right> =  \frac{\gla^2}{s}
e^{-2
\la^{-4} s^2} \SP \delta^{(6)}(p_1 + p_2 - p_3 -
p_4) ,
\lll{wrong amp}
\eea

\noi
where $\left| I \right> = \left| \phi_1 \phi_2 \right>$, $\left| F \right> = \left| \phi_3 \phi_4 \right>$, and
$s$ is the invariant mass-squared of $\left| \phi_1 \phi_2 \right>$.
The cross section is cutoff-independent only if the $T$ matrix is cutoff-independent.
Thus \reff{wrong amp} shows that the cutoff appears in physical quantities if noncanonical operators are
neglected.  If we were to use an infinite cutoff, then the $\OR(\gla^2)$ contribution to the cross section would
be
cutoff-independent but higher-order contributions would be infinite.

To test whether our procedure for computing $\M$ leads to cutoff-independent physical quantities, we now include
noncanonical operators.  This means that we have to include the first term in \reff{T exp}:

\bea
\left< \phi_3 \phi_4 \right|  H_I(\la)  \left| \phi_1 \phi_2 \right> &=& \frac{1}{p_1^+ + p_2^+}
e^{-\la^{-4} \bx{2}{FI}} \gla^2 \left< \phi_3 \phi_4 \right|  V^{(2)}(\la)  \left| \phi_1 \phi_2 \right>_1 \n
&=& \frac{\gla^2}{s} \left( 1 - e^{-2 \la^{-4} s^2} \right) \SP \delta^{(5)}(p_1 + p_2 - p_3 - p_4) ,
\eea

\noi
where 
the result for the annihilation contribution to the second-order reduced interaction is used, and we use 
the fact that we only need the on-energy-shell part.
This contribution shifts the $T$ matrix so that the total $T$ matrix is

\bea
\delta(P_F^- - P_I^-) \left< \phi_3 \phi_4 \right| T(P_I^-) \left| \phi_1 \phi_2 \right> =  \frac{\gla^2}{s} \SP
\delta^{(6)}(p_1 + p_2 - p_3 -
p_4) .
\eea

\noi
Since $\gla^2$ is cutoff-independent at second-order [$d\gla^2/d \la = \OR(\gla^4)$], this 
is the correct cutoff-independent result to $\OR(\gla^2)$.

Our restrictions on the Hamiltonian have led to its matrix elements being such that they compensate for the 
presence of the cutoff in physical quantities.  No proof exists, but we expect that our procedure leads to
a renormalized Hamiltonian that produces exactly correct perturbative scattering amplitudes order-by-order.
When this Hamiltonian is used nonperturbatively, however, there should be some cutoff dependence, as we have 
mentioned in previous sections.

\section{$\left< \phi_2 \phi_3 \right| \M \left| \phi_1 \right>$ to Third Order}

We wish to calculate $\left< \phi_2 \phi_3 \right| \M \left| \phi_1 \right>$ to third
order to demonstrate how a higher-order calculation proceeds and to derive the running coupling in our approach.
In Appendix D, we deduce that matrix elements of even-order
reduced interactions cannot have contributions from three interacting particles.  This means that
$\left< \phi_2 \phi_3 \right| \M \left| \phi_1 \right>$ has only odd-order contributions.
The first-order contribution is defined in terms of the first-order reduced interaction, which we have defined
in \reff{O1}.  The third-order contribution is defined in terms of the third-order reduced interaction.
In this section, we compute the third-order contribution,
neglecting the cutoff-independent part of the third-order reduced interaction.
\reff{CI cc soln} shows that a calculation of the
cutoff-independent part would be a fifth-order calculation, 
which we prefer to avoid at this time.

According to Eq. (\ref{eq:cc soln}), the cutoff-dependent part of the 
third-order reduced interaction is given by 

\bea
\Vthreeonetotwo = \left[ \dVthreeonetotwo - B_{1,2}
\left<\phi_2 \phi_3 \right| V^{(1)} \left| \phi_1 \right>\right]_\lz .
\lll{third}
\eea

\noi
$\dVthreeonetotwo$ is defined by

\bea
\dVthreeonetotwo &=& \frac{1}{2} \sum_K 
\left<\phi_2 \phi_3 \right| V^{(1)} \left| K \right> \left<K \right| V^{(2)}(\la') \left| \phi_1 \right> 
T_2^{(\la,\la')} \n
&+& \frac{1}{2} \sum_K \left<\phi_2 \phi_3 \right| V^{(2)}(\la') \left| K \right> 
\left<K \right| V^{(1)} \left| \phi_1 \right> T_2^{(\la,\la')}
\n
&+& \frac{1}{4} \sum_{K,L}  \left<\phi_2 \phi_3 \right| V^{(1)} \left| K \right> \left<K \right| V^{(1)} \left| L
\right>
\left<L \right| V^{(1)} \left| \phi_1 \right> T_3^{(\la,\la')} ,
\lll{to}
\eea

\noi
where we are suppressing the dependence of the cutoff functions on the states.  Note that 
$B_{1,2}=C_3$ and is given by

\bea
B_{1,2} &=& C_3 = \left[ \SP p_1^+ \delta^{(5)}(p_1 - p_2 - p_3) \right]^{-1} \dVthreeonetotwo
\:\rule[-4mm]{.1mm}{9mm}_{\; \vec p_{2 \perp} = \vec p_{3 \perp} = 0 \, ; \, 
p_2^+ = p_3^+ = \frac{1}{2} p_1^+} \n
&=& \left[ \left<\phi_2 \phi_3 \right| V^{(1)} \left| \phi_1 \right> \right]^{-1}
\dVthreeonetotwo
\:\rule[-4mm]{.1mm}{9mm}_{\; \vec p_{2 \perp} = \vec p_{3 \perp} = 0 \, ; \, 
p_2^+ = p_3^+ = \frac{1}{2} p_1^+} .
\lll{C3 def}
\eea

\noi
\reff{third} then becomes

\bea
\Vthreeonetotwo = \left[ \dVthreeonetotwo - \dVthreeonetotwo
\:\rule[-4mm]{.1mm}{9mm}_{\; \vec p_{2 \perp} = \vec p_{3 \perp} = 0 \, ; \, 
p_2^+ = p_3^+ = \frac{1}{2} p_1^+} \right]_\lz \hspace{-.3in} .
\lll{third2}
\eea

\noi
This indicates that $\left<\phi_2 \phi_3 \right| V^{(3)}_{\CD}(\la) \left| \phi_1 \right>$ can be
computed solely in terms of $\dVthreeonetotwo$, which we now calculate.

All the matrix elements of the second-order reduced interaction 
that can appear on the RHS of \reff{to} were calculated in the previous
section.  When $\dVthreeonetotwo$ is expanded in terms of the possible intermediate states and possible 
contributions to matrix elements of $V^{(1)}$ and $V^{(2)}(\la')$, we see that it has contributions from a number of
different processes.  We define $\dVthreeonetotwo_i$ to be the contribution to $\dVthreeonetotwo$ from 
the process shown in Fig. 4-i, where $1 \leq i \leq 15$.
\begin{figure}
\centerline{\epsffile{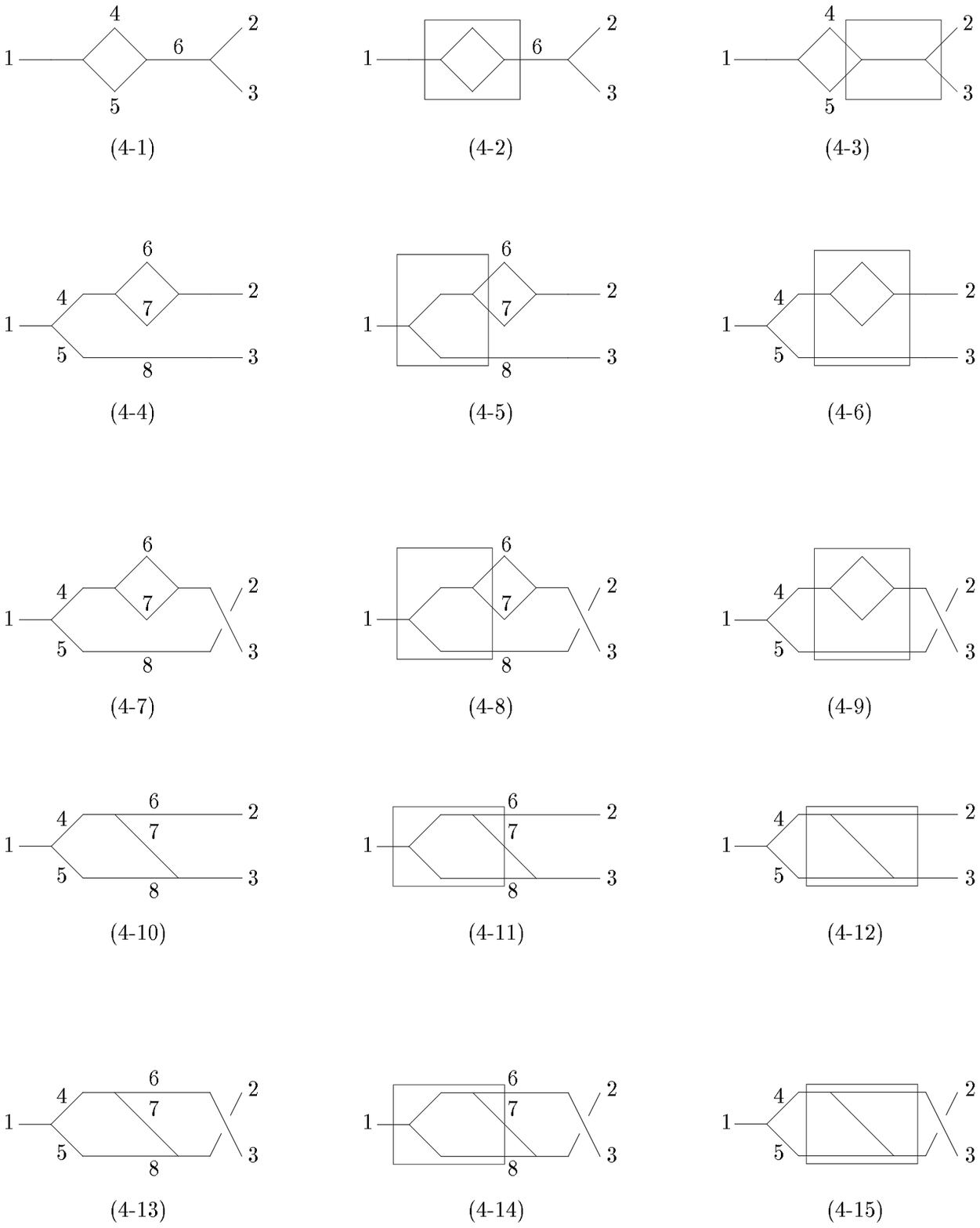}}
\caption{A diagrammatic representation of the contributions to $\dVthreeonetotwo$.  The numbers label the
particles, the
bare vertices represent matrix elements of $V^{(1)}$, and the boxes represent contributions to matrix elements
of $V^{(2)}(\la')$ corresponding to the processes depicted in the boxes.}
\end{figure}  
In Fig. 4, the plain vertices represent matrix elements of $V^{(1)}$ and the boxes represent
contributions to matrix elements of $V^{(2)}(\la')$ corresponding to the processes depicted in the boxes.
For example, the box in Fig. 4-12 represents 
$\left<\phi_2 \phi_3 \right| V^{(2)}(\la') \left| \phi_4 \phi_5 \right>_6$
[see Fig. 2-6 and \reff{exchange}].
Note that not every contribution to every matrix element of $V^{(2)}(\la')$ that appears in \reff{to}
appears in Fig 4.  
For example, $\left<\phi_2 \phi_3 \right| V^{(2)}(\la') \left| \phi_4 \phi_5 \right>_9$ 
[see Fig. 2-9 and \reff{other exchange}] does not appear in Fig. 4, because when it is paired with $\left<\phi_4 \phi_5
\right| V^{(1)} \left| \phi_1 \right>$ and
we integrate over $p_4$ and $p_5$, it gives an identical contribution to 
$\left<\phi_2 \phi_3 \right| V^{(2)}(\la') \left| \phi_4 \phi_5 \right>_6$.  
For this reason, $\dVthreeonetotwo_{12}$, shown in Fig. 4-12, includes contributions from both.  Then
$\dVthreeonetotwo$ 
is

\bea
\dVthreeonetotwo = \sum_{i=1}^{15} \dVthreeonetotwo_i .
\lll{R}
\eea

For each contribution to $\dVthreeonetotwo$, we use the change of variables

\bea
p_2 &=& (y p_1^+, y \vec p_{1 \perp} + \vec q_\perp) , \n
p_3 &=& ([1-y] p_1^+, [1-y] \vec p_{1 \perp} - \vec q_\perp) ,
\eea

\noi
which leads to the initial and final free masses

\bea
M_I^2 &=& 0 , \n
M_F^2 &=& \frac{\qsq}{y (1-y)} .
\lll{external}
\eea

\subsection{The One-Particle Self-Energy Contributions}

The contributions to the matrix element of the third-order change to the reduced interaction 
from one-particle self-energies, shown in Figs. 4-1, 4-2, and 4-3, 
can be combined because they have similar structures.  For each of them we use

\bea
\left| L \right> &=& \left| \phi_4 \phi_5 \right> , \n
\left| K \right> &=& \left| \phi_6 \right> , \n
\eea

\noi
and the change of variables

\bea
p_4 &=& (x p_1^+, x \vec p_{1 \perp} + \vec r_\perp) , \n
p_5 &=& ([1-x] p_1^+, [1-x] \vec p_{1 \perp} - \vec r_\perp) ,
\eea

\noi
which leads to the intermediate free masses

\bea
M_K^2 &=& 0 , \n
M_L^2 &=& \frac{\rsq}{x (1-x)} ,
\eea

\noi
and $p_6=p_1$.  The contribution from Fig. 4-1 is

\bea
\dVthreeonetotwo_1 &=& \frac{1}{8} \int D_4 D_5 D_6 \left<\phi_2 \phi_3 \right| V^{(1)} \left| \phi_6 \right> 
\left<\phi_6 \right| V^{(1)} \left| \phi_4 \phi_5 \right>
\left<\phi_4 \phi_5 \right| V^{(1)} \left| \phi_1 \right> T_3^{(\la,\la')}(F,K,L,I) \n
&=& \frac{N}{2 \pi^2} \int d^4 r_\perp dx \frac{1}{x (1-x)} T_3^{(\la,\la')}(F,K,L,I) ,
\lll{hold r1}
\eea

\noi
where

\bea
N = \frac{\pi^2}{4} p_1^+ \delta^{(5)}(p_1 - p_2 - p_3) .
\eea

\noi
The contribution from Fig. 4-2 is

\bea
\dVthreeonetotwo_2 &=& \frac{1}{2} \int D_6 
\left<\phi_2 \phi_3 \right| V^{(1)} \left| \phi_6 \right> \left<\phi_6 \right| V^{(2)}(\la') \left| \phi_1
\right>
T_2^{(\la,\la')}(F,K,I) \n
&=& \frac{N}{12} \sqrt{\frac{\pi}{2}} \la'^2 T_2^{(\la,\la')}(F,K,I) ,
\eea

\noi
and the contribution from Fig. 4-3 is

\bea
&&\dVthreeonetotwo_3 = \frac{1}{4} \int D_4 D_5 
\left<\phi_2 \phi_3 \right| V^{(2)}(\la') \left| \phi_4 \phi_5 \right>_1 \left<\phi_4 \phi_5 \right| V^{(1)}
\left|
\phi_1 \right> 
T_2^{(\la,\la')}(F,L,I) \n
&=& \frac{N}{2 \pi^2} \int d^4 r_\perp dx \frac{1}{x (1-x)} \left( \frac{1}{M_L^2} + \frac{1}{M_F^2} \right)
\left( 1 - e^{-2 \la'^{-4} M_L^2 M_F^2} \right) T_2^{(\la,\la')}(F,L,I) .
\eea

It is easiest to combine these three contributions
before doing the integral in \reff{hold r1}.  This leads to

\bea
\sum_{i=1}^3 \dVthreeonetotwo_i &=& \frac{N}{12} \int_0^\infty d(M_L^2) \left\{ e^{-2 \la'^{-4} M_L^4} \left[
\frac{1}{M_L^2}
+ \frac{M_F^2}{M_L^4} + \frac{2}{M_F^2} \right. \right. \n
&+& \left. \left. e^{2 \la'^{-4} M_L^2 M_F^2} \left( \frac{1}{\Bx{}{FL}} - \frac{1}{M_L^2}
+ \frac{M_L^2}{M_F^2 \Bx{}{FL}} - \frac{1}{M_F^2} \right) \right] - [\la' \rightarrow \la] \right\} .
\eea

\noi
This one-dimensional integral cannot be done analytically for arbitrary $M_F^2$; so we consider it to be
a function of $M_F^2$ that must be computed numerically.

\subsection{The Two-Particle Self-Energy Contributions}

The contributions to the matrix element of the third-order change to the reduced interaction from two-particle
self-energies, shown in Figs. 4-4, 4-5, and 4-6, can be combined.  For each of these we use

\bea
\left| L \right> &=& \left| \phi_4 \phi_5 \right> , \n
\left| K \right> &=& \left| \phi_6 \phi_7 \phi_8\right> , \n
\eea

\noi
and the change of variables

\bea
p_6 &=& (x p_2^+, x \vec p_{2 \perp} + \vec r_\perp) , \n
p_7 &=& ([1-x] p_2^+, [1-x] \vec p_{2 \perp} - \vec r_\perp) ,
\eea

\noi
which leads to the intermediate free masses

\bea
M_L^2 &=& \frac{\qsq}{y (1-y)} , \n
M_K^2 &=& \frac{\qsq}{y (1-y)} + \frac{\rsq}{x (1-x) y} ,
\eea

\noi
and

\bea
p_4 &=& p_2 , \n
p_5 &=& p_8 \; \, = \; \, p_3 .
\eea

\noi
Then these contributions are given by

\bea
\dVthreeonetotwo_4 &=& \frac{N}{2 \pi^2} \int d^4 r_\perp dx \frac{1}{x (1-x) y^2} T_3^{(\la,\la')}(F,K,L,I) , \n
\dVthreeonetotwo_5 &=& \frac{N}{2 \pi^2} \int d^4 r_\perp dx \frac{1}{x (1-x) y^2} \left( \frac{1}{\Bx{}{KL}} -
\frac{1}{M_L^2} 
\right) \left( 1 - e^{2 \la'^{-4} \bx{}{KL} M_L^2} \right) T_2^{(\la,\la')}(F,K,I) , \n
\dVthreeonetotwo_6 &=& \frac{N}{12} \sqrt{\frac{\pi}{2}} \la'^2 T_2^{(\la,\la')}(F,L,I) .
\eea

\noi
Writing their sum as an integral leads to

\bea
&&\sum_{i=4}^6 \dVthreeonetotwo_i = \frac{N}{12} \int_0^\infty d(\Bx{}{KL}) \left\{ e^{-2 \la'^{-4} \bx{2}{KL}} 
\left[ \frac{1}{\Bx{}{KL}}
- \frac{M_F^2}{\Bx{2}{KL}} - \frac{2}{M_F^2} \right. \right. \n
&+& \left. \left. e^{- 2 \la'^{-4} \bx{}{KL} M_F^2} \left( - \frac{1}{\Bx{}{KL} + M_F^2} - \frac{1}{\Bx{}{KL}}
+ \frac{\Bx{}{KL}}{M_F^2 (\Bx{}{KL}+M_F^2)} + \frac{1}{M_F^2} \right) \right] - [\la' \rightarrow \la] \right\}
\n
&=& \sum_{i=1}^3 \dVthreeonetotwo_i \:\rule[-4mm]{.1mm}{8mm}_{\; M_F^2 \rightarrow -M_F^2} .
\eea

From Fig. 4,

\bea
\sum_{i=7}^9 \dVthreeonetotwo_i &=& \sum_{i=4}^6 \dVthreeonetotwo_i \:\rule[-4mm]{.1mm}{8mm}_{\; 2
\leftrightarrow 3} ,
\eea

\noi
but since $\sum_{i=4}^6 \dVthreeonetotwo_i$ is a function only of $M_F^2$ and $N$, which are invariant under 
$2 \leftrightarrow 3$,

\bea
\sum_{i=7}^9 \dVthreeonetotwo_i &=& \sum_{i=4}^6 \dVthreeonetotwo_i .
\eea

\subsection{The Exchange Contributions}

For the contributions to the matrix element of the third-order change to the reduced interaction from exchange
processes, shown in Figs. 4-10, 4-11, and 4-12, we use

\bea
\left| L \right> &=& \left| \phi_4 \phi_5 \right> , \n
\left| K \right> &=& \left| \phi_6 \phi_7 \phi_8\right> ,
\eea

\noi
and the change of variables

\bea
p_4 &=& (x p_1^+, x \vec p_{1 \perp} + \vec r_\perp) , \n
p_5 &=& ([1-x] p_1^+, [1-x] \vec p_{1 \perp} - \vec r_\perp) ,
\eea

\noi
which leads to the intermediate free masses

\bea
M_L^2 &=& \frac{\rsq}{x (1-x)} , \n
M_K^2 &=& \frac{\qsq}{y} + \frac{\rsq}{(1-x)} + \frac{(\vec r_\perp - \vec q_\perp)^2}{x-y} ,
\eea

\noi
and

\bea
p_6 &=& p_2 , \n
p_7 &=& ([x-y] p_1^+, [x-y] \vec p_{1 \perp} + [\vec r_\perp - \vec q_\perp]) , \n
p_8 &=& p_5 .
\eea

\noi
Then 

\bea
&&\sum_{i=10}^{12} \dVthreeonetotwo_i = \frac{N}{\pi^2} \int d^4 r_\perp dx \frac{1}{x (1-x) (x-y)} \n
&\times& \left\{ \left[ \left( \frac{1}{M_L^2}
- \frac{1}{\Bx{}{KL}} \right) \left( \frac{1}{M_K^2} - \frac{1}{\Bx{}{FK}} \right) e^{2 \la'^{-4} \bx{}{FK}
M_K^2}
\right. \right. \n
&+& \left. \left.  \left( \frac{1}{\Bx{}{KL}} - 
\frac{1}{M_L^2} \right) \frac{M_F^2 - 2 M_K^2}{\Bx{}{FK} M_K^2 + \Bx{}{KL} M_L^2}
e^{2 \la'^{-4} ( \bx{}{FK} M_K^2 + \bx{}{KL} M_L^2 )} \right. \right. \n
&+& \left. \left. \left( \frac{1}{\Bx{}{KL}} - \frac{1}{\Bx{}{FK}} \right) \left( \frac{1}{M_L^2} - 
\frac{1}{\Bx{}{FL}} \right) e^{2 \la'^{-4} \bx{}{FL} M_L^2} \right. \right. \n
&+& \left. \left. \left( \frac{1}{\Bx{}{FK}} - \frac{1}{\Bx{}{KL}} \right) 
\frac{M_F^2 - 2 M_L^2}{\Bx{}{FK} M_K^2 + \Bx{}{KL} M_L^2}
e^{2 \la'^{-4} (\bx{}{FK} M_K^2 + \bx{}{KL} M_L^2)} \right] - [\la' \rightarrow \la] \right\} .
\lll{5d}
\eea

\noi
This five-dimensional integral also cannot be done analytically.  From Fig. 4,

\bea
\sum_{i=13}^{15} \dVthreeonetotwo_i = \sum_{i=10}^{12} \dVthreeonetotwo_i \:\rule[-4mm]{.1mm}{8mm}_{\; 2 
\leftrightarrow 3} = \sum_{i=10}^{12} \dVthreeonetotwo_i 
\:\rule[-4mm]{.1mm}{8mm}_{\; y \rightarrow 1-y \, ; \, \vec q_\perp \rightarrow - \vec q_\perp} \hspace{-.3in}.
\eea

\subsection{The Running Coupling}

We now have the complete matrix element of the third-order change to the reduced interaction 
in terms of numerical functions of $\vec q_\perp$, the transverse momentum of particle 2 in the
center-of-mass frame, and $y$, the fraction of the total longitudinal momentum carried by particle
2.  We first use the matrix element to compute the scale dependence of $\gla$.  According \reff{C3 def},
we need the $\vec p_{2 \perp} = \vec p_{3 \perp} =0$ limit of $\dVthreeonetotwo$. 
For both $\vec p_{2 \perp}$
and $\vec p_{3 \perp}$ to be zero, $\vec q_\perp$ must be zero, in which case $M_F^2$ is also zero.  
In this limit,
all the integrals in $\dVthreeonetotwo$ can be done analytically, 
although care must be used because there are delicate cancellations
of divergences among the different parts of each integrand.  Note that the $\vec q_\perp=0$ limit
of $\dVthreeonetotwo$ has no dependence on $p_2^+$ or $p_3^+$.  The result of applying \reff{C3 def} is

\bea
B_{1,2} = C_3 = \frac{3}{512 \pi^3} \log \frac{\la'^2}{\la^2},
\eea

\noi
which from \reff{scale dep} implies

\bea
\gla = \glap + \frac{3}{512 \pi^3} \glap^3 \log \frac{\la'^2}{\la^2} + \OR(\glap^5).
\lll{rc}
\eea

\noi
This equation tells us how the coupling is related at two different scales.  
Since $\la' > \la$, \reff{rc} shows that $\gla > \glap$.  This means the coupling grows as
we reduce the amount by which free masses can change.  This shows that the theory is asymptotically free, 
as expected.  

In conventional covariant perturbation theory with the minimal subtraction (MS) renormalization scheme, 
one obtains, for the scale dependence of the coupling [17],

\bea
g_\mu = g_{\mu'} + \frac{3}{512 \pi^3} g_{\mu'}^3 \log \frac{\mu'^2}{\mu^2} + \OR(g_{\mu'}^5),
\eea

\noi
where $\mu$ and $\mu'$ are two different renormalization points.  We see that
our coupling changes with our cutoff in the same manner that the MS coupling
changes with the renormalization point.  This is what one expects due to the scheme independence of
the one-loop beta function.  However, since there is no simple connection between our scheme and the 
MS scheme, we expect there to be no simple relation between the respective couplings at higher orders.

\subsection{Numerical Results for the Three-Point Matrix Element}

We now proceed to calculate the matrix element of the cutoff-dependent part of the third-order reduced
interaction.
From \reff{third2} and the fact $\vec p_{2 \perp} = \vec p_{3 \perp} = 0$ implies that $\vec q_\perp = 0$,

\bea
\left<\phi_2 \phi_3 \right| V^{(3)}_{\CD}(\la) \left| \phi_1 \right> = \left[ \dVthreeonetotwo - 
\dVthreeonetotwo\:\rule[-4mm]{.1mm}{8mm}_{\; \vec q_\perp \rightarrow 0} \right]_\lz ,
\lll{NM part}
\eea

\noi
where we use the fact that the $\vec q_\perp=0$ limit of $\dVthreeonetotwo$ is independent of $p_2^+$ and
$p_3^+$.  
Since the Hamiltonian is invariant under transverse rotations and $\vec q_\perp$ is the only transverse vector 
on  which $\dVthreeonetotwo$ depends, \reff{NM part} indicates that 
$\dVthreeonetotwo$ is a function of $\vec q_\perp^{\:2}$.  By inspection, we can see that
$\dVthreeonetotwo$ is the product of $N$ and some dimensionless quantity, and is 
the difference of a function of $\la$ and the same function with $\la \rightarrow \la'$.
Also, by our assumption of approximate transverse locality, the RHS of \reff{NM part} 
can be expanded in powers of $\vec q_\perp$.  Thus

\bea
\dVthreeonetotwo - \dVthreeonetotwo \:\rule[-4mm]{.1mm}{8mm}_{\; \vec q_\perp \rightarrow 0} = 
N \sum_{i=1}^\infty f_i(y) \vec q_\perp^{\, 2 i} 
\left( \la^{-2 i} - \la'^{-2 i} \right) ,
\lll{R exp}
\eea

\noi
where the $f_i$'s are unknown dimensionless functions of $y$.  This implies

\bea
\left<\phi_2 \phi_3 \right| V^{(3)}_{\CD}(\la) \left| \phi_1 \right> &=& \left[ \dVthreeonetotwo - 
\dVthreeonetotwo\:\rule[-4mm]{.1mm}{8mm}_{\; \vec q_\perp \rightarrow 0} \right]_\lz \n
&=& N \sum_{i=1}^\infty f_i(y) \vec q_\perp^{\, 2 i} \la^{-2 i} \n
&=& \left( \dVthreeonetotwo - \dVthreeonetotwo \:\rule[-4mm]{.1mm}{8mm}_{\; \vec q_\perp \rightarrow 0} \right) 
\:\rule[-5mm]{.1mm}{11mm}_{\; \la' \rightarrow \infty} .
\lll{O3}
\eea

\noi
Using this equation and our result for $\dVthreeonetotwo$, it is straightforward to write 
$\left< \phi_2 \phi_3
\right| V^{(3)}_{\CD}(\la) \left| \phi_1 \right>$ as a numerical function of $\vec q_\perp$, $y$, and $\la$.

By inspection, $\dVthreeonetotwo$ is symmetric under $y \rightarrow 1-y$ and each 
$\dVthreeonetotwo_i$
for $i \le 9$ is a function only of $M_F^2$, the free mass of the final state.  
It is not clear from inspection if the rest of $\dVthreeonetotwo$ 
is a function only of $M_F^2$ and there is no reason that it should be.  This means that the matrix element of
the Hamiltonian is not necessarily a function only of $M_F^2$.

Our result for the three-point matrix element of the Hamiltonian through third order, neglecting any
cutoff-independent contribution to the third-order reduced interaction, is

\bea
\left< \phi_2 \phi_3 \right| \M \left| \phi_1 \right> = e^{-\la^{-4} M_F^4} \left[ \D{1}{2 3} \gla + \gla^3
\left<\phi_2 \phi_3 \right| V^{(3)}_{\CD}(\la) \left| \phi_1 \right> \right] .
\eea

\noi
To get an idea of the size of the noncanonical contribution to the matrix element, we can compare it to the 
canonical contribution.  To do this, we need to choose a value for the coupling.  We would like to choose a
large coupling so that we can get a pessimistic estimate of whether or not the expansion for the Hamiltonian
is converging.  When the coupling is large, the second term in \reff{rc} is nearly as large as the
first.  The value of $\log(\la^{\prime 2}/\la^2)$ is our choice, but the natural value is 1, because then the range 
of scales over which off-diagonal
matrix elements of the Hamiltonian are being removed is comparable to the range of scales that remain.
Thus we estimate that a large coupling is $\gla^2 = 512 \pi^3/3$.  

Using this coupling, 
Fig. 5 compares the canonical and noncanonical contributions to the matrix element as a function of the
free mass of the final state.  The plot is the result of a numerical
computation using the VEGAS Monte Carlo algorithm for multidimensional integration [18].
The contributions to the matrix element are plotted in units of the matrix element of the 
unregulated canonical
interaction, and we consider only the situation in which  the two final-state particles share the total 
longitudinal
momentum equally.  The solid curve represents the canonical contribution  and the diamonds
represent the noncanonical contribution.   The statistical error bars from the Monte Carlo integration are too
small 
to be visible.

Fig. 5 shows that for the case we consider, the noncanonical corrections
to the matrix element of the Hamiltonian are small compared to the canonical part, even when the perturbative 
expansion of the coupling is breaking down. 
However, this does not necessarily imply that corrections to physical quantities from the noncanonical
part of this matrix element would be small.  Investigation of this would require a fifth-order calculation in
this
theory; so we feel that this is not worth investigating until we consider QCD.
\begin{figure}
\centerline{\epsffile{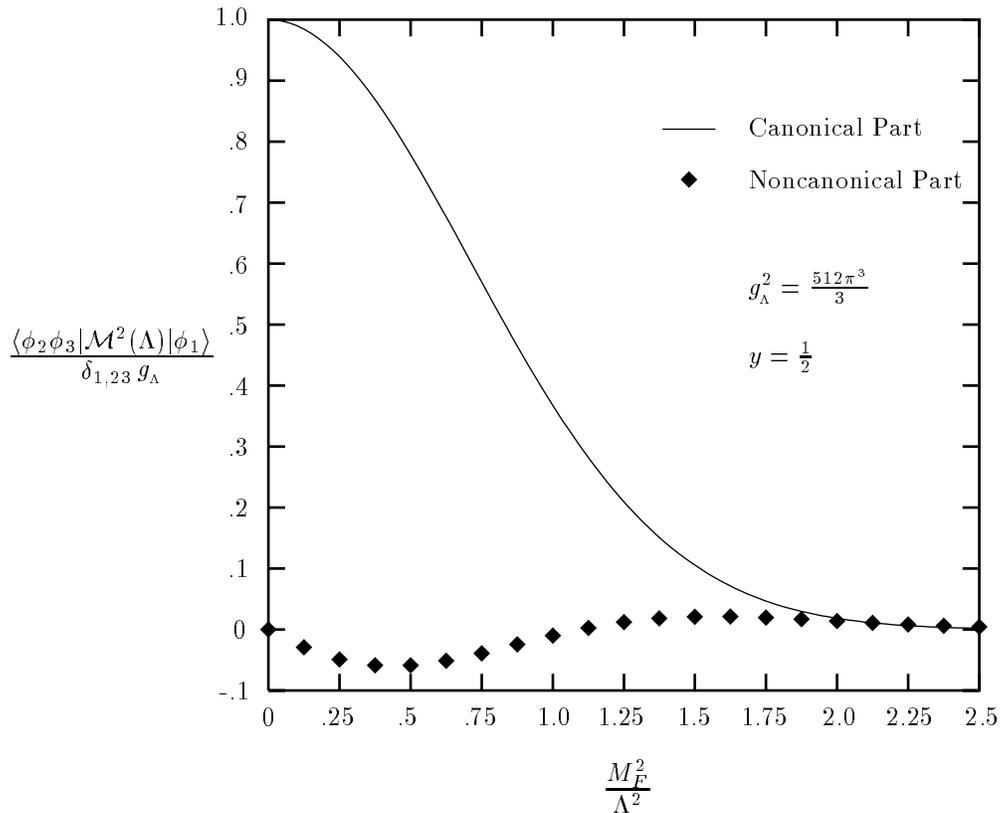}}
\caption{A comparison of the canonical and noncanonical contributions to 
$\left< \phi_2 \phi_3 \right| \M \left| \phi_1 \right>$ as a function of the
free mass of the final state.  The contributions to the matrix element are plotted in units of the 
matrix element of the unregulated canonical
interaction, and we consider only the situation in which  the two final-state particles share the total
longitudinal
momentum equally.  The coupling used in the plot is $\gla^2 = 512 \pi^3/3$.  
The solid curve represents the canonical contribution and the diamonds
represent the noncanonical contribution. The statistical error bars from the Monte Carlo integration are too
small 
to be visible.}
\end{figure}

For the same coupling, 
Fig. 6 shows the noncanonical
part of the matrix element as a function of
the fraction of the total longitudinal momentum carried by particle 2 for fixed free mass of the final
state.  It is plotted in units of the 
matrix element of the unregulated canonical interaction for
three different values of the free mass of the final state.
\begin{figure}
\centerline{\epsffile{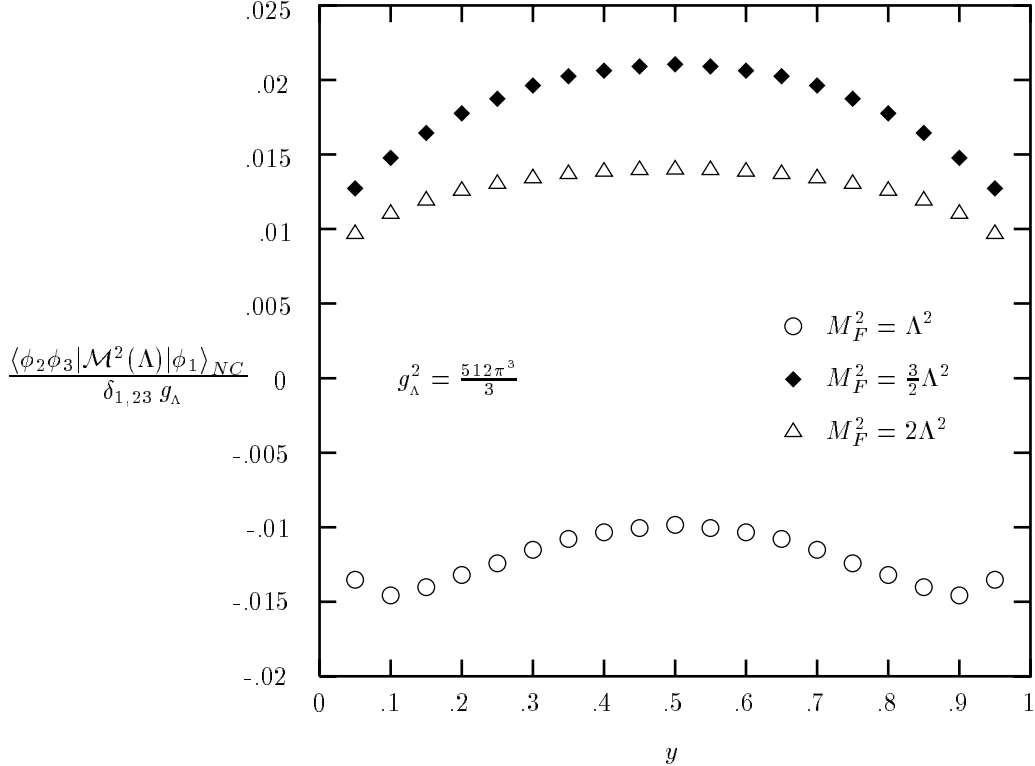}}
\caption{The noncanonical part of $\left< \phi_2 \phi_3 \right| \M \left| \phi_1 \right>$
as a function of
the fraction of the total longitudinal momentum carried by particle 2 for fixed free mass of the final
state.  It is plotted in units of the 
matrix element of the unregulated canonical interaction for
three different values of the free mass of the final state.  The coupling used in the plot is 
$\gla^2 = 512 \pi^3/3$.  The statistical  error bars from the Monte Carlo
integration are too small to be visible.}
\end{figure}
This plot shows that the noncanonical part of the matrix element is a function of $y$ for fixed 
$M_F^2$, and thus is not a function only of $M_F^2$.

\section{Conclusions}

We have outlined three conditions under which hadron states will rapidly converge in an expansion in free-particle 
Fock-space sectors in a
HLFQCD approach.  First, the diagonal matrix elements of the Hamiltonian in 
the free-particle Fock-space basis must be dominated by the free part of the Hamiltonian.  
Second, the off-diagonal matrix elements of the Hamiltonian in this basis must quickly decrease as
the difference of the free masses of the states increases.   Third, the mass of a free
state must quickly increase as the number of particles in the state increases.
We have argued that the Hamiltonian cannot meet these conditions unless
the vacuum is dominated by few-body free sectors; so we force the vacuum to be empty by
requiring every particle to have a positive longitudinal momentum.
We have argued that if the vacuum is empty and the Hamiltonian can be derived perturbatively, 
then the first and third conditions on the Hamiltonian should
be automatically satisfied in HLFQCD, and the second condition can be enforced by suppressing the Hamiltonian's 
matrix elements between states that differ in free mass by more than a cutoff. 

The cutoff we use, like all regulators in HLFFT, violates a number of physical principles of the theory.
This means that the Hamiltonian cannot be  just the
canonical Hamiltonian with masses and couplings redefined by renormalization.  Instead, the Hamiltonian must
be allowed to contain all operators that are consistent with the unviolated physical principles of the 
theory.  We have shown that if we require
the Hamiltonian to produce cutoff-independent physical quantities and we require it to respect the unviolated
physical principles of the theory, then the matrix elements of the Hamiltonian are uniquely determined in terms 
of the fundamental parameters of the theory.

To force the Hamiltonian to produce cutoff-independent physical quantities, we require it to be
unitarily equivalent
to itself at a slightly larger cutoff.  To make the Hamiltonian respect the unviolated physical principles of the
theory, we place a number of additional constraints on it. First, the Hamiltonian must conserve momentum and be
invariant under boosts and transverse rotations.  Second, it should respect approximate cluster decomposition.  
Third, it should be approximately local in the transverse directions.  Fourth, it should reproduce the correct 
free field theory in the noninteracting limit.  Fifth, it
can depend only on the fundamental parameters of the theory and the cutoff.  
Sixth, the Hamiltonian must produce the correct
second-order scattering amplitudes.  The seventh and final restriction we place on
the Hamiltonian is that it must lead to perturbative scattering amplitudes with the correct dependence on the coupling.

If the physical principles that we use are consistent with the physical principles 
violated  by the cutoff, such as
gauge invariance and Lorentz invariance, then the matrix elements of the Hamiltonian that follow from our
procedure
must lead to physical quantities that respect all the physical principles of the theory.  

Our key assumption is that in an  asymptotically free theory the matrix elements of the Hamiltonian can be
computed perturbatively,  provided the Hamiltonian is defined at a scale where the couplings are sufficiently
small.  This assumption and the rapid convergence of hadron states in the expansion in free-particle 
Fock-space sectors will allow
us to approximate physical quantities in HLFQCD with a finite number of degrees of freedom.

As an illustration of our method for calculating the Hamiltonian,  we have calculated some of its second- and
third-order matrix elements in $\phi^3$ theory in six dimensions.  We have shown that the second-order matrix
elements naturally obey the restrictions we place on the Hamiltonian and that they remove the cutoff dependence from
physical quantities. We have also shown how the coupling depends on the scale in our approach and how asymptotic
freedom arises. Finally, we have demonstrated that the third-order noncanonical corrections to the
$\phi_1 \rightarrow \phi_2 \phi_3$ matrix element of the Hamiltonian that arise from the cutoff-dependent part
of the reduced interactions
are small, even when the perturbative expansion of the coupling is beginning to break down.

We intend to apply our method to QCD; however, direct extension of the method to light-front QCD is
complicated by  the presence of quark masses.  This is because quark masses are additional fundamental 
parameters on which the Hamiltonian's matrix elements can depend.  This complicates the method
for
determining the matrix elements, and thus the extension of our method to include particle masses requires 
further research.  

On the other hand, if quark masses are unimportant, then the only fundamental parameter in QCD is the quark-gluon
coupling.   It is then quite straightforward to apply our method to determine the Hamiltonian. There are two
main complications that are absent in scalar theory.   The first is algebraic complexity due to the presence of
spin, color, and multiple canonical vertices.  The  second is the need for an additional cutoff to regulate
divergent contributions to interactions involving gluons with vanishing longitudinal momenta.  This cutoff can
be implemented on the canonical interactions at the very beginning of the calculation of the Hamiltonian and
can be taken to its limit at the end of a calculation of a physical quantity.   This second cutoff does not
complicate our method for determining the Hamiltonian, but having to take the cutoff to its limit significantly
complicates the process of diagonalizing the Hamiltonian. One of the requirements that we have placed on the method 
presented here is that it will lead to physical quantities that are finite and well-behaved in this limit.  A
demonstration of this quality is part of the natural next step, application of our methods to a nonperturbative
relativistic HLFQCD calculation.

\newpage

\begin{tabbing}
{\large \bf APPENDIX A:}$\;$\={\large \bf Particle-Number Dependence of Free-Hamiltonian}\\
\>{\large \bf Eigenvalues}
\end{tabbing}

\vspace{0.1in}

The free Hamiltonian satisfies the Fock-space eigenvalue equation

\bea
\Mf \left| F \right> = M_F^2 \left| F \right> ,
\eea

\noi
where $\left| F \right>$ is any free-particle Fock-space state.  We define $N_F$ to be the
number of particles in $\left| F \right>$.  The eigenvalue $M_F^2$ depends on $N_F$:

\bea
M_F^2 = \sum_{i=1}^{N_F} \frac{\vec r_{i \perp}^{\,2}}{x_i} ,
\eea

\noi
where $x_i$ is the fraction of the total longitudinal momentum of $\left| F \right>$ that is carried by 
particle $i$ and $\vec r_{i \perp}$ is
the particle's transverse momentum in $\left| F \right>$'s center-of-mass frame.  
Momentum conservation implies that

\bea
\sum_{i=1}^{N_F} x_i = 1 .
\lll{mom}
\eea

\noi
If $N_F$ is increased, the $x_i$'s get smaller to maintain \reff{mom}.  Let us assume that

\bea
x_i \sim \frac{1}{N_F} ,
\lll{worst}
\eea

\noi
in order to maintain \reff{mom}.  Let us also assume that $\vec r_{i \perp}^{\,2}$ is non-negligible and
approximately independent of $i$.  Then

\bea
M_F^2 \sim \sum_{i=1}^{N_F} N_F = N_F^2 .
\eea

\noi
If the assumption in \reff{worst} does not hold, then $M_F^2$ increases even faster with $N_F$. \footnote{Note that in 
equal-time field
theory the free energy of a state increases only linearly with the number of particles, assuming they
have non-negligible momenta.}

This argument fails if many particles have negligible center-of-mass transverse momenta. However,
this is unlikely in a confining theory such as QCD because confinement suppresses any state containing particles
with large transverse separation from the other particles in the state, and thus favors states containing
particles
with some non-negligible center-of-mass transverse momenta.

\newpage
\begin{tabbing}
{\large \bf APPENDIX B:}$\;$\={\large \bf Canonical Light-Front Massless Scalar Field Theory in Six}\\
\>{\large \bf Dimensions}
\end{tabbing}

\vspace{.1in}
The purpose of this appendix is to state our conventions.  A more complete discussion of canonical light-front
scalar field theory can be found in Ref. [19].

With our conventions, any six-vector $a$ is written in the form

\bea
a = (a^+, a^-, \vec a_\perp) ,
\lll{six vector}
\eea

\noi
where in terms of equal-time vector components,

\bea
a^{\pm} = a^0 \pm a^5
\eea

\noi
and

\bea
\vec a_\perp = \sum_{i=1}^4 a_\perp^i \hat e_i = \sum_{i=1}^4 a^i \hat e_i,
\eea

\noi
where $\hat e_i$ is the unit vector pointing in the $i^{\hspace{.1mm}\mathrm{th}}$ direction.  
The inner product is

\bea
a \cdot b = \frac{1}{2} a^+ b^- + \frac{1}{2} a^- b^+ - \vec a_\perp \cdot \vec b_\perp ,
\eea

\noi
and

\bea
\vec a_\perp^2 = \vec a_\perp \cdot \vec a_\perp .
\eea

A spacetime coordinate is a six-vector, and according to \reff{six vector}, it is written

\bea
x = (x^+, x^-, \vec x_\perp) .
\eea

\noi
The time component is chosen to be $x^+$.  Here $x^-$ is referred to as the longitudinal component, and $\vec x_\perp$
contains the transverse components.

The gradient operator is treated just like any other six-vector.  Its components are

\bea
\partial^{\pm} = 2 \frac{\partial}{\partial x^\mp} 
\eea

\noi
and

\bea
\partial_\perp^i = - \frac{\partial}{\partial x_\perp^i} .
\eea

The canonical Lagrangian density for massless scalar field theory with a three-point interaction is

\bea
{\cal L} = \frac{1}{2} \partial \phi \cdot \partial \phi - \frac{g}{3!} \phi^3 ,
\eea

\noi
where the gradient operator is understood to act on the first spacetime-dependent function to its right,
unless otherwise indicated by parentheses.  The canonical Hamiltonian follows from ${\cal L}$ and the
assumption that the field vanishes at spacetime infinity.  It is

\bea
H = \int d^4 x_\perp dx^- \left[ - \frac{1}{2} \phi \vec \partial_\perp^{\,2} \phi + \frac{g}{3!} \phi^3 \right]
.
\eea

To this point, $\phi$ has been regarded as a classical field.  We work in the Schr\"odinger representation, 
where operators are time-independent and states are time-dependent.  Thus we quantize the field by
defining it as a time-independent function of free-particle creation and annihilation operators:

\bea
\phi(x^-,\vec x_\perp) = \int D_1 \left[ a_1 e^{-i p_1 \cdot x} + a_1^\dagger e^{i p_1 \cdot x} \right]
\rule[-3mm]{.1mm}{7mm}_{\;
x^+ = 0},
\eea

\noi
where

\bea
D_i \equiv \frac{d^4 p_{i \perp} dp_i^+}{64 \pi^5 p_i^+} 
\eea

\noi
and 

\bea
p_i^+ \ge 0 .
\eea

$p^+$ and $\vec p_\perp$ are the conjugate momenta to $x^-$ and $\vec x_\perp$; so they are
referred to as the longitudinal and transverse momenta, respectively.  The creation and annihilation operators
follow the convention

\bea
a_i = a(p_i^+, \vec p_{i \perp}) ,
\eea

\noi
and have the commutation relations

\bea
[a_i , a_j^\dagger] &=& \D{i}{j} 
\eea

\noi
and

\bea
[a_i , a_j] = [a_i^\dagger , a_j^\dagger] = 0 ,
\eea

\noi
where

\bea
\D{abc\cdots}{a'b'c'\cdots} &=& \SP (p_a^+ + p_b^+ + p_c^+ + \cdots ) 
\delta^{(5)}(p_a + p_b + p_c + \cdots - p_{a'} - p_{b'} - p_{c'} - \cdots) 
\eea

\noi
and

\bea
\delta^{(5)}(p_i - p_j) &=& \delta(p_i^+ - p_j^+) \delta^{(4)}(\vec p_{i \perp} - \vec p_{j \perp}) .
\eea

Let $\p$ be the six-momentum operator and ${\cal M}$ be the invariant-mass operator. Since the momentum conjugate
to 
$x^+$ is $p^-$, the Hamiltonian is identified as $\p^-$, and it follows from

\bea
\p^2 = {\cal M}^2 
\eea

\noi
that

\bea
\p^- = \frac{\vec \p_\perp^{\,2} + {\cal M}^2}{\p^+} .
\lll{p minus}
\eea

The Hamiltonian can be written as the sum of a free part and an interacting part:

\bea
\p^- = H = h + v ,
\eea

\noi
where

\bea
h = \int D_1 \frac{\vec p_{1 \perp}^{\,2}}{p_1^+} a_1^\dagger a_1
\eea

\noi
and

\bea
v = (2 \pi)^5 g \int D_1 D_2 D_3 \left[ a_3^\dagger a_1 a_2 \delta^{(5)}(p_3 - p_1 - p_2) +
a_2^\dagger a_3^\dagger a_1 \delta^{(5)}(p_2 + p_3 - p_1) \right] .
\eea

\noi
The process of normal ordering $h$ and $v$ produces infinite constants that have no physical significance and
are dropped.  The Hamiltonian also contains operators that have nonzero matrix 
elements only if a particle can have a longitudinal momentum of zero.
These particle are dropped from the theory and thus the associated operators have no effect and are dropped.  We plan to 
extend our method at some point in the future by replacing
the effects of the dropped particles with interactions in the Hamiltonian.
Without these particles, the vacuum is empty.

The eigenstates of $h$ are

\bea
\left| \phi_1 \phi_2 \cdots \phi_n\right> = a_1^\dagger a_2^\dagger \cdots a_n^\dagger \left| 0 \right> ,
\eea

\noi
for any integer $n \ge 0$.  The associated eigenvalue equation is

\bea
h \left| \phi_1 \phi_2 \cdots \phi_n\right> = \sum_{i=1}^n p_i^- \left| \phi_1 \phi_2 \cdots 
\phi_n\right> ,
\eea

\noi
where

\bea
p_i^- = \frac{\vec p_{i \perp}^{\,2}}{p_i^+} 
\eea

\noi
and the sum is zero if $n=0$.

The noninteracting limit of \reff{p minus} is

\bea
h = \frac{\vec \p_\perp^{\,2} + {\cal M}_f^2}{\p^+} ,
\eea

\noi
where ${\cal M}_f$ is the free invariant-mass operator.  It has the eigenvalue equation

\bea
{\cal M}_f^2 \left| \phi_1 \phi_2 \cdots \phi_n\right> = M^2 \left| \phi_1 \phi_2 \cdots 
\phi_n\right> ,
\eea

\noi
where

\bea
M^2 = P^+ \sum_{i=1}^n p_i^- - \vec P_\perp^2 
\eea

\noi
and $P$ is the total momentum of the state.

Finally, in terms of the free states, the completeness relation is

\bea
{\bf 1} = \left| 0 \right> \left< 0 \right| + \int D_1 \left| \phi_1 \right> \left< \phi_1 \right| +
\frac{1}{2!} \int D_1 D_2 \left| \phi_1 \phi_2 \right> \left< \phi_1 \phi_2 \right| + \cdots .
\lll{completeness}
\eea

\newpage

\centerline{\large {\bf APPENDIX C: Proof of Unitarity of $U(\la,\la')$}}

\vspace{.1in}
The purpose of this appendix is to prove that $U(\la,\la')$, as defined by

\bea
\frac{d U(\la,\la')}{d (\la^{-4})} = T(\la) U(\la,\la') 
\lll{diff eq 2}
\eea

\noi
and

\bea
U(\la,\la) = {\bf 1} ,
\lll{ubc}
\eea

\noi
is unitary as long as $T(\la)$ is anti-Hermitian.  Assume that $T$ is anti-Hermitian.  
Multiply \reff{diff eq 2} on the left by $U^\dagger(\la,\la')$:

\bea
U^\dagger(\la,\la') \frac{d U(\la,\la')}{d (\la^{-4})} &=& U^\dagger(\la,\la') T(\la) U(\la,\la') \n
&=& - \left[ U^\dagger(\la,\la') T(\la) U(\la,\la') \right]^\dagger \n
&=& - \left[ U^\dagger(\la,\la') \frac{d U(\la,\la')}{d (\la^{-4})} \right]^\dagger \n
&=& - \frac{d U^\dagger(\la,\la')}{d (\la^{-4})} U(\la,\la') .
\eea

\noi
This is the same as

\bea
\frac{d}{d (\la^{-4})} \left[ U^\dagger(\la,\la') U(\la,\la') \right] = 0 ,
\lll{unitary diff 1}
\eea

\noi
which implies that $U^\dagger(\la,\la') U(\la,\la')$ is independent of $\la$.  Since
\reff{ubc} implies

\bea
U^\dagger(\la',\la') U(\la',\la') = {\bf 1} ,
\lll{unitary bc 1}
\eea

\noi
we conclude

\bea
U^\dagger(\la,\la') U(\la,\la') = {\bf 1} .
\lll{unitary cond 1}
\eea

Now multiply \reff{diff eq 2} on the right by $U^\dagger(\la,\la')$:

\bea
\frac{d U(\la,\la')}{d (\la^{-4})} U^\dagger(\la,\la') = T(\la) U(\la,\la') U^\dagger(\la,\la'),
\lll{diff eq 3}
\eea

\noi
and take the adjoint of this:

\bea
U(\la,\la') \frac{d U^\dagger(\la,\la')}{d (\la^{-4})}  = - U(\la,\la') U^\dagger(\la,\la') T(\la).
\lll{diff eq 4}
\eea

\noi
Adding Eqs. (\ref{eq:diff eq 3}) and (\ref{eq:diff eq 4}) gives

\bea
\frac{d}{d (\la^{-4})} \left( U(\la,\la') U^\dagger(\la,\la') \right) = \left[T(\la),U(\la,\la')
U^\dagger(\la,\la')
\right] .
\lll{unitary diff 2}
\eea

\noi
From \reff{ubc},

\bea
U(\la',\la') U^\dagger(\la',\la') = {\bf 1} .
\lll{unitary bc 2}
\eea

\noi
$U(\la,\la') U^\dagger(\la,\la')$ is a function of $\la$ that is uniquely determined by 
Eqs. (\ref{eq:unitary diff 2}) and (\ref{eq:unitary bc 2}).  Therefore, since the statement

\bea
U(\la,\la') U^\dagger(\la,\la') = {\bf 1}
\lll{unitary cond 2}
\eea

\noi
is a solution to Eqs. (\ref{eq:unitary diff 2}) and (\ref{eq:unitary bc 2}), it is a true statement.  
Since $U(\la,\la')$ satisfies
Eqs. (\ref{eq:unitary cond 1}) and (\ref{eq:unitary cond 2}), it is unitary.

\newpage
\appendix

\begin{tabbing}
{\large \bf APPENDIX D:}$\;$\={\large \bf Derivation of Reduced Interactions in Terms of Lower-Order}\\
\>{\large \bf Reduced Interactions}
\end{tabbing}

\vspace{.1in}

In Section 4, we derived a constraint on the $\OR(\gla^r)$ reduced interaction for $r \ge 1$:

\bea
&&V^{(r)}_{\CD}(\la) - 
V^{(r)}_{\CD}(\la') = \delta V^{(r)} - \sum_{s=2}^{r-1}  B_{r-s,s}
V^{(r-s)}(\la) .
\lll{CD coupled D}
\eea

\noi
Since we already know the first-order reduced interaction [see \reff{O1}], 
we wish to use this equation to compute the $\OR(\gla^r)$ reduced interaction for $r \ge 2$, in
terms of the lower-order reduced interactions.  

\vspace{.20in}

\noi
{\large \bf D.1  \hspace{.05in} The Cutoff-Dependent Part}
\vspace{.08in}

We begin by computing the cutoff-dependent part.  
Recall that momentum conservation implies that any matrix element 
of $V(\la)$ can be written as an expansion in unique products of
momentum-conserving delta functions.  This means that an arbitrary matrix element of 
\reff{CD coupled D} can be expanded in products of delta functions:

\bea
&&\sum_i \left<F \right| V^{(r)}_{\CD}(\la) \left| I \right>^{(i)} - 
\sum_i \left<F \right| V^{(r)}_{\CD}(\la') \left| I \right>^{(i)} \n
&=& \sum_i \left[ \dVome{r} - \sum_{s=2}^{r-1}  B_{r-s,s}
\left<F \right| V^{(r-s)}(\la) \left| I \right>\right]^{(i)} ,
\eea

\noi
where the $(i)$ superscripts denote that we are considering the $i^{\hspace{.1mm}\mathrm{th}}$ 
product of delta functions that can occur 
in a delta function expansion of $\left<F \right| V^{(r)}(\la) \left| I \right>$.  This
equation is equivalent to a set of equations, one for each possible product 
of delta functions:

\bea
\left<F \right| V^{(r)}_{\CD}(\la) \left| I \right>^{(i)} - 
\left<F \right| V^{(r)}_{\CD}(\la') \left| I \right>^{(i)} = \left[ \dVome{r} - \sum_{s=2}^{r-1}  B_{r-s,s}
\left<F \right| V^{(r-s)}(\la) \left| I \right>\right]^{(i)}  .
\lll{prod exp}
\eea

\noi
Cluster decomposition implies that we can write

\bea
\left<F \right| V^{(r)}_{\CD}(\la) \left| I \right>^{\,(i)} &=& 
\left[ \prod_{j=1}^{N_{\delta}^{(i)}} \delta_j^{(i)} \right] F^{(i)}_{\CD}(\{p_n\},\la),
\lll{gen delta prefactor}
\eea

\noi
where $\delta_j^{(i)}$ is the $j^{\hspace{.1mm}\mathrm{th}}$ delta function in the 
$i^{\hspace{.1mm}\mathrm{th}}$ product of delta functions (it includes the
longitudinal momentum factor), $N_{\delta}^{(i)}$ is the number of delta functions in the 
$i^{\hspace{.1mm}\mathrm{th}}$ product, and
$F^{(i)}_{\CD}(\{p_n\},\la)$ is a function of the cutoff and the momenta of the particles in 
the matrix element, but does not contain delta functions that fix momenta. 
We define $N_{\mathrm{part}}$ to be the number of particles in state $\left| I \right>$ plus the number of particles in
state $\left| F \right>$, and $n=1,2,3,\ldots,N_{\mathrm{part}}$.  Here $p_n$ is the momentum of particle $n$.
We define $N_{\mathrm{int}}^{(i)}$ to be the number of  interacting particles in
state $\left| I \right>$ plus the number of interacting particles in state $\left| F \right>$ for the $i^{\hspace{.1mm}\mathrm{th}}$
product of delta functions.  In order for the Hamiltonian to have the dimensions $(\mathrm{mass})^2$,
$F^{(i)}_{\CD}$ must have the dimensions $({\mathrm{mass}})^{6-2 N_{\mathrm{int}}^{(i)}}$.
Note that we are suppressing the dependence of the RHS of this equation on $r$.

We have assumed that any matrix element of the Hamiltonian can be expanded in powers of transverse momenta, not
including
the momentum-conserving delta functions; so

\bea
F^{(i)}_{\CD}(\{p_n\},\la) = \la^{6 - 2 N_{\mathrm{int}}^{(i)}} \sum_{\left\{ m_{nt} \right\}}
z^{\left\{m_{nt} 
\right\}}_i
\Big( \{p_n^+\}\Big) \prod_{n=1}^{N_{\mathrm{part}}} \prod_{t=1}^4 \left( \frac{ p_{n \perp}^t }{\la}
\right)^{m_{nt}} ,
{\lll F}
\eea

\noi
where $t=1,2,3,4$ denotes a component of
transverse momentum and  $m_{nt}$ is a non-negative integer index associated with transverse momentum component
$t$ of particle $n$.  The sum is over all values of each of the $m_{nt}$'s,
subject to the constraint that

\bea
6 - 2 N_{\mathrm{int}}^{(i)} - \sum_{n,t} m_{nt} \neq 0 ,
\lll{nonmarg cond}
\eea

\noi
which is necessary to avoid terms in the momentum expansion that are cutoff-independent.
The $z^{\left\{m_{nt} \right\}}_i$'s are the coefficients for the momentum expansion.  
They depend on the $m_{nt}$'s and are functions of the longitudinal momenta of the particles.

Since the RHS of \reff{prod exp} has the same product of delta functions as the LHS, we can write

\bea
&&\left[ \dVome{r} - \sum_{s=2}^{r-1}  B_{r-s,s}
\left<F \right| V^{(r-s)}(\la) \left| I \right>\right]^{(i)} = \left[ \prod_{j=1}^{N_{\delta}^{(i)}}
\delta_j^{(i)} \right]  
G^{(i)}(\{p_n\},\la,\la'),
\lll{G prefactor}
\eea

\noi
where $G^{(i)}$ has dimensions $({\mathrm{mass}})^{6 - 2 N_{\mathrm{int}}^{(i)}}$, 
and by inspection of the LHS of \reff{prod exp} and \reff{gen delta prefactor} is a function of the momenta 
of the particles, $\la$, and $\la'$.  Substitution of Eqs. (\ref{eq:G prefactor}) and (\ref{eq:gen delta
prefactor}) into \reff{prod exp} yields

\bea
F^{(i)}_{\CD}(\{p_n\},\la) - F^{(i)}_{\CD}(\{p_n\},\la') = G^{(i)}(\{p_n\},\la,\la') ,
\lll{eq}
\eea

\noi
where the momenta in this equation are constrained by the delta function conditions.

Since the LHS of \reff{eq} is the difference of a function of $\la$ and the
same function with $\la \rightarrow \la'$, $G^{(i)}$ must be as well.  Since the 
LHS of \reff{eq} can be expanded in powers of transverse momenta, $G^{(i)}$ 
must have the form

\bea
G^{(i)}(\{p_n\},\la,\la') &=& \sum_{\left\{ m_{nt} \right\}} Z_i^{\left\{ m_{nt} \right\}}
\Big( \{p_n^+\}\Big) \n
&\times& \left[\la^{6 - 2 N^{(i)}_{\mathrm{int}}} \prod_{n,t} \left( \frac{ p_{n \perp}^t}{\la} 
\right)^{m_{nt}} - \la'^{\, 6 - 2 N^{(i)}_{\mathrm{int}}} \prod_{n,t} \left( \frac{ p_{n \perp}^t }{\la'}
\right)^{m_{nt}}\right] ,
\lll{G}
\eea

\noi
where the sum is restricted by \reff{nonmarg cond}.

Substituting Eqs. (\ref{eq:G}) and (\ref{eq:F}) into \reff{eq}, we find

\bea
&&\la^{6 - 2 N^{(i)}_{\mathrm{int}}} \sum_{\left\{ m_{nt} \right\}} z_i^{\left\{m_{nt} 
\right\}}
\Big( \{p_n^+\}\Big) \prod_{n,t} \left( \frac{ p_{n \perp}^t }{\la}
\right)^{m_{nt}} \n
&-& \la'^{\, 6 - 2 N^{(i)}_{\mathrm{int}}} \sum_{\left\{ m_{nt} \right\}} z_i^{\left\{m_{nt} 
\right\}}
\Big( \{p_n^+\}\Big) \prod_{n,t} \left( \frac{ p_{n \perp}^t}{\la'}
\right)^{m_{nt}} \n
&=& \sum_{\left\{ m_{nt} \right\}} Z_i^{\left\{ m_{nt} \right\}}
\Big( \{p_n^+\}\Big) \left[\la^{6 - 2 N^{(i)}_{\mathrm{int}}} \prod_{n,t} \left( \frac{ p_{n \perp}^t}{\la} 
\right)^{m_{nt}} - \la'^{\, 6 - 2 N^{(i)}_{\mathrm{int}}} \prod_{n,t} \left( \frac{ p_{n \perp}^t }{\la'}
\right)^{m_{nt}} \right] .
\lll{expand match}
\eea
 
\noi
Matching powers of transverse momenta on both sides of this equation gives

\bea
&&z_i^{\left\{ m_{nt} \right\}}
\Big( \{p_n^+\}\Big) \left[ \la^{6 - 2 N^{(i)}_{\mathrm{int}} - \Sigma_{n,t} \, m_{nt}}  - 
\la'^{\, 6 - 2 N^{(i)}_{\mathrm{int}} - \Sigma_{n,t} \, m_{nt}} \right] \n
&=& Z_i^{\left\{ m_{nt} \right\}}
\Big( \{p_n^+\}\Big) \left[ \la^{6 - 2 N^{(i)}_{\mathrm{int}} - \Sigma_{n,t} \, m_{nt}}  - 
\la'^{\, 6 - 2 N^{(i)}_{\mathrm{int}} - \Sigma_{n,t} \, m_{nt}} \right] .
\lll{power match}
\eea

\noi
The factor in brackets cannot be zero because $\la \neq \la'$ and
\reff{nonmarg cond} holds.  Thus \reff{power match} implies

\bea
z_i^{\left\{ m_{nt} \right\}} &=& Z_i^{\left\{ m_{nt} \right\}} .
\lll{cc exp sol}
\eea

\noi
Then Eqs. (\ref{eq:F}), (\ref{eq:G}), and (\ref{eq:cc exp sol}) imply

\bea
F^{(i)}_{\CD}(\{p_n\},\la) = G^{(i)}(\{p_n\},\la,\la') \lzb ,
\lll{cc F}
\eea

\noi
where ``$\la \; \mathrm{terms}$" means that $G^{(i)}$ is to be expanded in powers of transverse
momenta and only the terms in the expansion that depend on $\la$ are to be kept.  From
Eqs. (\ref{eq:gen delta prefactor}), (\ref{eq:G prefactor}), and (\ref{eq:cc F}),

\bea
\left<F \right| V^{(r)}_{\CD}(\la) \left| I \right>^{\,(i)} = \left[ \dVome{r} - \sum_{s=2}^{r-1}  B_{r-s,s}
\left<F \right| V^{(r-s)}(\la) \left| I \right>\right]^{(i)}_\lz ,
\eea

\noi
where it is understood that the momentum-conserving delta functions are ignored for the purposes of
transverse-momentum expansions.
Since a matrix element is the sum of the contributions to it from different products of delta functions, both 
sides of this equation can be summed over $i$ to obtain

\bea
\left<F \right| V^{(r)}_{\CD}(\la) \left| I \right> = \left[ \dVome{r} - \sum_{s=2}^{r-1}  B_{r-s,s}
\left<F \right| V^{(r-s)}(\la) \left| I \right>\right]_\lz .
\lll{cc soln D}
\eea

\noi
This equation tells us how to calculate the cutoff-dependent part of the $\OR(\gla^r)$ reduced interaction 
in terms of lower-order contributions.  

\vspace{.20in}

\noi
{\large \bf D.2  \hspace{.05in} The Cutoff-Independent Part}

\vspace{.08in}

To complete the
solution, we need to specify how to compute the cutoff-independent part.  
It is useful to first consider which contributions to $V^{(r)}(\la)$ can be cutoff-independent.

A matrix element of the cutoff-independent part of $V^{(r)}(\la)$ can be expanded in products of delta
functions and in powers of transverse momenta just as was done for the cutoff-dependent part.  Thus we can write

\bea
\left<F \right| V^{(r)}_{\CI} \left| I \right> = \sum_i \left<F \right| V^{(r)}_{\CI} \left| I \right>^{\,(i)} ,
\eea

\noi
where

\bea
\left<F \right| V^{(r)}_{\CI} \left| I \right>^{\,(i)} = \left[ \prod_{j=1}^{N_{\delta}^{(i)}} \delta_j^{(i)}
\right]  F^{(i)}_{\CI}(\{p_n\})
\lll{CI prefactor}
\eea

\noi
and

\bea
F^{(i)}_{\CI}(\{p_n\}) = \la^{6 - 2 N^{(i)}_{\mathrm{int}}} \sum_{\left\{ m_{nt} \right\}}
w_i^{\left\{m_{nt} 
\right\}}
\Big( \{p_n^+\}\Big) \prod_{n=1}^{N_{\mathrm{part}}} \prod_{t=1}^4 \left( \frac{ p_{n \perp}^t }{\la}
\right)^{m_{nt}} ,
\lll{FCI}
\eea

\noi
where the sum is over all values of each $m_{nt}$, subject to the constraint

\bea
6 - 2 N_{\mathrm{int}}^{(i)} - \sum_{n,t} m_{nt} = 0 .
\lll{marg cond}
\eea

\noi
\reff{marg cond} ensures that all the terms in the expansion of $F^{(i)}_{\CI}$ are cutoff-independent.  

\reff{marg cond} places
constraints on the possible cutoff-independent contributions to the reduced interaction. 
Any contribution to a matrix element of $V^{(r)}(\la)$ has an $N^{(i)}_{\mathrm{int}} \ge 2$, but 
\reff{marg cond} can only hold if
$N_{\mathrm{int}}^{(i)} \le 3$.  Suppose that $N_{\mathrm{int}}^{(i)}=2$.  In this case,
$F^{(i)}_{\CI}(\{p_n\})$ can be written as a sum of terms, where each term corresponds to a distinct
pair of interacting particles and depends on their momenta and the total longitudinal momentum 
[see \reff{cluster 2} for an example]:

\bea
F^{(i)}_{\CI}(\{p_n\}) &=& \sum_m F^{(i,m)}_{\CI}(s_m, s_m', P^+) \n
&=& \sum_m F^{(i,m)}_{\CI}(s_m, P^+) ,
\eea

\noi
where $s_m$ and $s_m'$ are the momenta for the initial and final particles in the 
$m^{\hspace{.1mm}\mathrm{th}}$ interacting pair, 
and where we
have used the fact that for $N_{\mathrm{int}}^{(i)}=2$, momentum conservation implies $s_m=s_m'$.  According to
\reff{marg cond}, if $N^{(i)}_{\mathrm{int}}=2$, $F_{\CI}^{(i)}$ is cutoff-independent only if 
$\sum_{n,t} m_{nt} = 2$.  This means that either $F^{(i,m)}_{\CI}$ is quadratic in $\vec s_{m \perp}$ or it is zero.  
However, the matrix elements of the Hamiltonian are boost-invariant, as is the delta-function product in
\reff{CI prefactor}.  This means that $F^{(i)}_{\CI}$ must be boost-invariant, but it cannot be if 
$F^{(i,m)}_{\CI}$ is quadratic in $\vec s_{m \perp}$.  Thus $F^{(i,m)}_{\CI}=0$, $F^{(i)}_{\CI}=0$, and 
there are no cutoff-independent contributions to  the reduced interaction with two interacting particles.  

Note that contributions to the Hamiltonian
with two interacting particles are self-energies, and they change the particle dispersion relation.  
If they change the dispersion relation such that the coefficients of the free relation
become modified by interactions, then this can be viewed as
renormalization of the field operators, i.e. wave function renormalization.  
This effect is absent unless either $F^{(i)}_{\CI}$ or $F^{(i)}_{\CD}$ can be quadratic in
transverse momenta for $N^{(i)}_{\mathrm{int}}=2$.  We have just shown that this is not possible for
$F^{(i)}_{\CI}$, and according to \reff{nonmarg cond}, $F^{(i)}_{\CD}$ cannot be quadratic in
transverse momenta for $N^{(i)}_{\mathrm{int}}=2$; so there
is no wave function renormalization at any order in $\gla$ in our approach.

The only other possibility for cutoff-independent contributions to the reduced interaction is from three
interacting particles.  Recall that in Section 3 we placed a restriction on the Hamiltonian that stated that
each perturbative  scattering amplitude that it produces must depend only on odd powers of the coupling if the
number of particles changes by an odd number and only on even powers of the coupling if the number of particles
changes by an even number.  This implies that the $\OR(\gla^r)$ reduced interaction can change particle number
by any even number $s$ satisfying $0 \le s \le r$ if $r$ is even, or any odd number $s$ satisfying $1 \le s \le
r$ if $r$ is odd.  (We neglect to present the proof of this because it is tedious.\footnote{We assume that the
free-state matrix elements of the Hamiltonian can be chosen to be real.})  
This means that since $V^{(r)}_{\CI}$ must have three interacting particles, it can be nonzero only if
$r$ is odd.

According to \reff{marg cond}, if $N^{(i)}_{\mathrm{int}}=3$, then $F^{(i)}_{\CI}$ must be independent of all
transverse momenta.  Thus, if $r$ is even, then $V^{(r)}_{\CI}=0$, and if $r$ is odd,
then $V^{(r)}_{\CI}$ 
is the part of $V^{(r)}(\la)$ with three interacting particles and no transverse momentum dependence.
Since $V^{(1)}$ is cutoff-independent, it should satisfy these conditions and inspection of 
\reff{O1} shows that it does.  

Before proceeding with the calculation of $V^{(r)}_{\CI}$, we need to deduce a bit more about the 
relationship of $\gla$ to $\glap$.
According to \reff{running 1} and the surrounding discussion, 
this relationship is determined by the matrix element
$\left< \phi_2 \phi_3 \right| \delta V \left| \phi_1 \right>$, which can be expanded in powers
of $\glap$:

\bea
\left< \phi_2 \phi_3 \right| \delta V \left| \phi_1 \right> = \sum_{t=3}^\infty \glap^t 
\left< \phi_2 \phi_3 \right| \delta V^{(t)} \left| \phi_1 \right> .
\lll{last run}
\eea

\noi
Recall that $\delta V^{(t)}$ is built from products of $V^{(r)}(\la')$'s.   This implies that $\delta V^{(t)}$
can change particle number by 1 only if $t$ is odd, and thus \reff{last run} implies that the coupling runs at
odd orders; i.e. $C_s$ is zero if $s$ is even [see \reff{scale dep}].

To calculate the cutoff-independent part of the $\OR(\gla^r)$ reduced interaction, consider \reff{prod exp} with
$r \rightarrow r+2$:

\bea
&&\left<F \right| V^{(r+2)}_{\CD}(\la) \left| I \right>^{(i)} - 
\left<F \right| V^{(r+2)}_{\CD}(\la') \left| I \right>^{(i)} \n
&=& \left[ \dVome{r+2} - \sum_{s=2}^{r+1}  B_{r+2-s,s}
\left<F \right| V^{(r+2-s)}(\la) \left| I \right>\right]^{(i)} .
\lll{marg exp}
\eea

\noi
In order for $V^{(r)}(\la)$ to have a cutoff-independent part, $r$ must be odd.  We know $V^{(1)}$; so assume
$r \ge 3$.  To calculate the cutoff-independent
part of $V^{(r)}(\la)$, we only have to consider the case $N^{(i)}_{\mathrm{int}}=3$.  According to Eqs. 
(\ref{eq:gen delta prefactor}), (\ref{eq:F}), and
(\ref{eq:nonmarg cond}), if $N^{(i)}_{\mathrm{int}}=3$, then for any $t$,

\bea
\left<F \right| V^{(t)}_{\CD}(\la) \left| I \right>^{(i)}_{\mathrm{Ext. \;} \vec k_\perp \rightarrow 0} = 0 ,
\lll{limit}
\eea

\noi
where ``Ext. $\vec k_\perp \rightarrow 0$" means the limit in which the transverse momenta
in the external states are taken to zero.  Then the zero-external-transverse-momentum
limit of \reff{marg exp} is

\bea
&&0 = \left[ \dVome{r+2} \right]^{(i)}_{\mathrm{Ext. \;} \vec k_\perp \rightarrow 0} - \sum_{s=2}^{r+1}  
B_{r+2-s,s}
\left<F \right| V^{(r+2-s)}_{\CI} \left| I \right>^{\;(i)} ,
\eea

\noi
where the cutoff-dependent part of the matrix element in the sum is annihilated by the
limit and the cutoff-independent part is untouched.  If we take the first term in the sum
on the RHS and move it to the LHS, we get

\bea
&& B_{r,2}
\left<F \right| V^{(r)}_{\CI} \left| I \right>^{\;(i)} = \left[ \dVome{r+2} \right]^{(i)}_{\mathrm{Ext. \;} 
\vec k_\perp \rightarrow 0} - \sum_{s=3}^{r+1}  B_{r+2-s,s}
\left<F \right| V^{(r+2-s)}_{\CI} \left| I \right>^{\;(i)} .
\eea

\noi
We can sum over $i$, keeping in mind that we are restricted to the case of
three interacting particles.  Then we find

\bea
\left<F \right| V^{(r)}_{\CI} \left| I \right> &=& \frac{1}{B_{r,2}} \left[ \dVome{r+2} 
\:\rule[-4mm]{.1mm}{8mm}_{\; \mathrm{Ext. \;} \vec k_\perp \rightarrow 0} - \sum_{s=3}^{r+1}  B_{r+2-s,s}
\left<F \right| V^{(r+2-s)}_{\CI} \left| I \right> \right]_{\mathrm{3 \; int. \; 
part.}} \hspace{-.5in} ,
\lll{CI cc soln D}
\eea

\noi
where ``$\mathrm{3 \; int. \; part.}$" means the 
products of momentum-conserving delta functions that appear on the
RHS are to be examined and only the contributions from three interacting particles 
are to be kept.

This equation tells us how to compute the cutoff-independent part of the  $\OR(\gla^r)$ reduced interaction for odd
$r \ge 3$  in terms of the cutoff-dependent part and lower-order reduced interactions.   It is an integral
equation\footnote{It is very difficult to prove that integral equations of this type have a unique solution; so
we simply assume it is true in this case.} because $V_{\CI}^{(r)}$ is nested with lower-order reduced
interactions in integrals in $\delta V^{(r+2)}$.  
A cursory examination of the definition of $\delta V^{(r+2)}$
may lead  one to believe that \reff{CI cc soln D} is useless because it requires one to know $V^{(r+1)}(\la')$. 
However,  since $r$ is odd, $V^{(r+1)}(\la')=V^{(r+1)}_{\CD}(\la')$, and the dependence of $\delta V^{(r+2)}$ on
$V^{(r+1)}_{\CD}(\la')$  can be replaced with further dependence on $V^{(r)}_{\CI}$ and $V^{(r)}_{\CD}(\la')$ 
(and lower-order reduced interactions) using \reff{cc soln D} with $r \rightarrow r+1$.

\newpage

\noi
{\large \bf Acknowledgments}

We would like to thank Billy D. Jones, David G. Robertson, Stanis{\l}aw D. G{\l}azek, S\'{e}rgio Szpigel, and Ken
Wilson for many
useful discussions about renormalization and light-front field theory. 
We also thank R. J. Furnstahl, Roger D. Kylin, and Matthias Burkardt for their useful comments
on the text.  This work has been partially supported by National Science Foundation grant
PHY-9511923.

\hspace{.3in}


\end{document}